\documentclass[usenatbib]{mnras}
\usepackage{lscape}
\usepackage{amsmath,amssymb}
\usepackage{mathrsfs}	
\usepackage{multicol}
\usepackage[dvipdfmx]{graphicx}
\usepackage{textcomp}
\usepackage{txfonts}
\usepackage[T1]{fontenc}
\usepackage{aecompl}

\catcode`\@=11
\def\gta{\ifmmode{\,\mathrel{\mathpalette\@versim>\,}}
    \else{$\,\mathrel{\mathpalette\@versim>}\,$}\fi}
\def\lta{\ifmmode{\,\mathrel{\mathpalette\@versim<\,}}
    \else{$\,\mathrel{\mathpalette\@versim<}\,$}\fi}
\def\@versim#1#2{\lower 2.9truept \vbox{\baselineskip 0pt \lineskip
    0.5truept \ialign{$\m@th#1\hfil##\hfil$\crcr#2\crcr\sim\crcr}}}
\catcode`\@=12

\makeatletter
\def\blfootnote{\xdef\@thefnmark{}\@footnotetext}
\makeatother

\newcommand{\Mdm}{M_{\rm dm}}
\newcommand{\kms}{\,{\rm km\,s}^{-1}}
\newcommand{\Mfst}{M_{0,\star}}
\newcommand{\Jfst}{J_{0,\star}}
\newcommand{\rfst}{r_{0,\star}}
\newcommand{\vfst}{v_{0,\star}}

\newcommand{\fJPst}{f({\bf J})}
\newcommand{\gJ}{g({\bf J})}
\newcommand{\hJ}{h({\bf J})}
\newcommand{\kJ}{k({\bf J})}
\newcommand{\tJ}{T({\bf J})}

\newcommand{\Mfdm}{M_{0,\rm dm}}
\newcommand{\Jfdm}{J_{0,\rm dm}}
\newcommand{\rfdm}{r_{0,\rm dm}}
\newcommand{\vfdm}{v_{0,\rm dm}}
\newcommand{\Jtdm}{J_{\rm t, dm}}
\newcommand{\Jcdm}{J_{\rm c, dm}}

\newcommand{\Mftil}{\tilde{M}_{0,\rm dm}}

\newcommand{\Jftil}{\tilde{J}_{0,\rm dm}}
\newcommand{\Jctil}{\tilde{J}_{\rm c, dm}}
\newcommand{\Jttil}{\tilde{J}_{\rm t, dm}}
\newcommand{\xib}{\boldsymbol \xi}

\newcommand{\Mstot}{M_{\rm tot, \star}}
\newcommand{\Mdmtot}{M_{\rm tot, dm}}
\newcommand{\Mvir}{M_{200}}
\newcommand{\rvir}{r_{200}}
\newcommand{\rs}{r_{\rm s}}
\newcommand{\rc}{r_{\rm c}}
\newcommand{\rt}{r_{\rm t}}
\newcommand{\rh}{r_{\rm h}}
\newcommand{\rst}{\tilde{r}_{\rm s}}
\newcommand{\rct}{\tilde{r}_{\rm c}}
\newcommand{\rtt}{\tilde{r}_{\rm t}}
\newcommand{\Mdmt}{\tilde{M}_{\rm tot, dm}}
\newcommand{\Mdn}{M_{\rm dyn}}
\newcommand{\Reff}{R_{\rm e}}

\newcommand{\MratRe}{(M_{\text{dm}}/M_{\star})|_{R_{\rm e}}}
\newcommand{\Mrat}{(M_{\text{dm}}/M_{\star})|_{3 {\rm kpc}}}

\newcommand{\Phidm}{\Phi_{\text{dm}}}
\newcommand{\rhodm}{\rho_{\rm dm}}
\newcommand{\fJ}{f_{\star}({\bf J})}
\newcommand{\fJh}{f_{\star}(h)}

\newcommand{\fJdm}{f_{\rm dm}({\bf J})}
\newcommand{\fJdmh}{f_{\rm dm}(h)}
\newcommand{\Phist}{\Phi_{\star}}
\newcommand{\rhost}{\rho_{\star}}
\newcommand{\Phitot}{\Phi_{\text{tot}}}

\newcommand{\nobs}{n_{\star}^{\text{obs}}}
\newcommand{\nobsi}{n_{\star, i}^{\text{obs}}}
\newcommand{\vlos}{v_{\rm los}}
\newcommand{\vlosk}{v_{{\rm los}, k}}

\newcommand{\Ltot}{\mathcal{L}}
\newcommand{\Ln}{\mathcal{L}_n}
\newcommand{\Lv}{\mathcal{L}_v}

\newcommand{\nmod}{n_{\star}}

\begin{document}

\date{Revised Draft, 1 June 2018}
\title[Action-based dynamical models of dwarf spheroidal galaxies: application to Fornax]
{Action-based dynamical models of dwarf spheroidal galaxies: application to Fornax}
\author[]{Raffaele Pascale$^{1,2}$\thanks{E-mail: raffaele.pascale2@unibo.it}, Lorenzo Posti$^{3}$, Carlo Nipoti$^{1}$, James Binney$^{4}$
\\ \\
$^{1}$Dipartimento di Fisica e Astronomia, Universit\`a di Bologna, via Piero Gobetti 93/2, I-40129 Bologna, Italy \\
$^{2}$INAF Osservatorio Astronomico di Bologna, Via Piero Gobetti 93/3, I-40129 Bologna, Italy \\
$^{3}$Kapteyn Astronomical Institute, University of Groningen, P.O. Box 800, 9700 AV Groningen, The Netherlands \\
$^{4}$Rudolf Peierls Centre for Theoretical Physics, Keble Road, Oxford, OX1 3NP, United Kingdom}

\maketitle

\begin{abstract}
We present new dynamical models of dwarf spheroidal galaxies (dSphs) in which both the 
stellar component and the dark halo are described by analytic distribution functions 
that depend on the action integrals. In their most general form these distribution
functions can represent axisymmetric and possibly rotating stellar systems. Here, as
a first application, we model the Fornax dSph, limiting ourselves, for simplicity, 
to the non rotating, spherical case. The models are compared with state-of-the-art 
spectroscopic and photometric observations of Fornax, exploiting the knowledge 
of the line-of-sight velocity distribution of the models and accounting for the 
foreground contamination from the Milky Way. The model that best fits the structural
and kinematic properties of Fornax has a cored dark halo, with core size $\rc\simeq1.03$ 
kpc. The dark-to-luminous mass ratio is $(\Mdm/M_{\star})|_{\Reff}\simeq9.6$
within the effective radius $\Reff \simeq 0.62\,$kpc and $(\Mdm/M_{\star})|_{3 {\rm kpc}} 
\simeq 144$ within 3 kpc. The stellar velocity distribution is isotropic almost over
the full radial range covered by the spectroscopic data and slightly radially anisotropic
in the outskirts of the stellar distribution. The dark-matter annihilation $J$-factor and
decay $D$-factor are, respectively, $\log_{10}(J$ $[$GeV$^2$ cm$^{-5}])\simeq18.34$ and 
$\log_{10}(D$ $[$GeV cm$^{-2}])\simeq18.55$, for integration angle $\theta = 0.5^{\circ}$.
This cored halo model of Fornax is preferred, with high statistical significance, to both
models with a Navarro, Frenk and White dark halo and simple mass-follows-light models. 

\end{abstract}
\begin{keywords}
dark matter - galaxies: dwarf - galaxies: individual: Fornax - galaxies: kinematics and dynamics - galaxies: structure
\end{keywords}

\section{Introduction}
\label{sec:int}

The dwarf spheroidal galaxies (dSphs) are gas poor faint stellar systems with roughly 
elliptical shape. Due to their very low surface brightness, dSphs are observed only in
the local Universe, but similar galaxies are expected to be ubiquitous in the cosmos. 
The nearest and best known dSphs belong to the Local Group, being satellites of the
Milky Way (hereafter MW) and M31. dSphs are interesting astrophysical targets for 
several reasons. In the standard $\Lambda$ cold dark matter ($\Lambda$CDM) cosmological 
model, dwarf galaxies are the building blocks of more massive galaxies, so the 
knowledge of their properties is a fundamental step in understanding galaxy formation.
Moreover, there is now much evidence (essentially based on measures of the stellar 
line-of-sight velocities; \citealt{Aaronson1986}, \citealt{Battaglia2013}) that these 
galaxies are hosted in massive and extended dark halos, which usually dominate the stellar
components even in the central parts. dSphs almost completely lack emission in bands other 
than the optical, so they are natural locations at which to look for high-energy signals
from annihilating or decaying dark-matter particles (e.g. \citealt{Evans2016}). These facts
make dSphs ideal laboratories in which to study dark matter, to understand the processes
that drive galaxy formation and to test cosmology on the smallest scales, where there is
potential tension between the observational data and the predictions of the $\Lambda$CDM 
model (\citealt{Bullock2017}).

The core/cusp problem is a clear example of this controversy: on the one hand, 
cosmological dark-matter only $N$-body simulations predict cuspy dark halo density
profiles; on the other hand, the rotation curves of low surface brightness disc 
and gas-rich dwarf galaxies favour shallower or cored dark-matter density 
distributions (\citealt{deBlok2009} and references therein). Also for dSphs, for 
which the determination of the dark-matter density distribution is more difficult, 
there are indications that cored dark-matter density profiles may be favoured with 
respect to cuspy profiles (\citealt{Kleyna2003}, \citealt{Goerdt2006},
\citealt{Battaglia2008}, \citealt{Walker2011}, \citealt{Salucci2012}, \citealt{Amorisco2013},
\citealt{Zhu2016}), though this finding is still debated (\citealt{Richardson2014}, \citealt{Strigari2017}).
It must be stressed, however, that cored dark halos in dSphs do not necessarily
imply a failure of $\Lambda$CDM: DM-only cosmological simulations may not reliably
predict the present-day dark-matter distribution in dSphs because, by definition, 
they neglect the effects of baryons on the dark halos. Even in a galaxy that is 
everywhere dark-matter dominated today, baryons must have been locally dominant 
in the past to permit star formation. Therefore, the effect of baryon physics on 
the dark halo is expected to be important also in dSphs. For instance, 
\cite{NipotiBinney2015} showed how, due to the fragmentation of a disc in cuspy 
dark halo, dynamical friction may cause the halo to flatten the original cusp
into a core even before the formation of the first stars (see also \citealt{ElZant2001}, 
\citealt{MoMao2004}, \citealt{Goerdt2010}, \citealt{Cole2011}, \citealt{ArcaSedda2017}).
Moreover, the results of hydrodynamical simulations suggest that, following star formation, 
supernova feedback can also help to flatten the central dark-matter distribution, 
by expelling the gas (\citealt{Navarro1996}, \citealt{Read2005}) and thus inducing 
rapid fluctuations in the gravitational potential (\citealt{Mashchenko2006}, 
\citealt{Pontzen2012},  \citealt{Tollet2016}).

The determination of the dark-matter distribution in observed dSphs relies on the 
combination of high-quality observational data and sophisticated dynamical modelling
(see \citealt{Battaglia2013} for a review). With the advent of the latest generation
of spectrographs and thanks to wide-field surveys, today we have relatively large
samples of individual stars in dSphs with measured line-of-sight velocities, allowing, 
in principle, for a detailed study of the dynamics of these nearby dwarf galaxies. 
To exploit this kind of information optimally, much effort has gone into
developing reliable, physical and self-consistent techniques for modelling galaxies 
(\citealt{Strigari2008}, \citealt{Walker2009}, \citealt{AmoriscoEvans2011}, 
\citealt{Jardel2012}, \citealt{BreddelsHelmi2013}). However, the process of 
understanding the properties of the dark halos of dSphs is far from complete.

If the effects of the tidal field of the host galaxy (for instance the MW) are
negligible, a dSph can be modeled as a collisionless equilibrium stellar system, 
which is completely described in terms of time-independent distribution functions
(hereafter DFs). In this work we present a novel mass modelling method for dSphs 
based on DFs depending on the action integrals ${\bf J}$. The actions are integrals
of motion that can be complemented by canonically conjugate (angle) variables to 
form a set of phase-space canonical coordinates. The action $J_i$  is
\begin{equation}
 J_i = \frac{1}{2\pi}\oint_{\gamma_i} {\bf p}\cdot \text{d}{\bf q},
\end{equation}
where ${\bf p}$ and ${\bf q}$ are any canonical phase-space coordinates and
$\gamma_i$ is a closed path over which the corresponding angle conjugate to
$J_i$ makes a full oscillation.  Actions are ideal labels for stellar orbits,
and an action-based DF specifies how the galaxy's orbits are populated.
\cite{Binney2014} proved that spherical galaxy models based on $f({\bf J})$
DFs depending on actions can easily be extended to systems with rotation
and flattening. Moreover, actions are adiabatic invariants (i.e. they are
unchanged under slow changes in the potential). This property makes the
$f(\bf J)$ models particularly suitable to model multi-component galaxies, in
which some components may have grown adiabatically. For instance, during the
accumulation of the baryonic component in a dark halo, the total
gravitational potential changes, and so does the halo's density distribution.
However, if the halo responds adiabatically, the distribution of its
particles in action space remains unchanged. 

Regardless of whether a galaxy is really assembled by adiabatic addition of 
components, one can readily assign each component a likely action-based DF that 
completely specifies the component's mass and angular momentum, and then quickly 
solve for the gravitational potential that all components jointly generate
(\citealt{Piffl2015}). Once that is done, it is easy to compute any observable
whatsoever.  Thanks to all of these features, dynamical models relying on 
action-based DFs have proved successful in modelling the MW (\citealt{Piffl2015},
\citealt{Binney2015}, \citealt{Sanders2015}, \citealt{ColeBinney2017}). 

The application of the $f({\bf J})$ models to dSphs is also very promising,
because it exploits the possibility of computing physical models with known DFs,
given large kinematic samples of line-of-sight velocity measures (see 
\citealt{Williams2015}, \citealt{Jeffreson2017}). In particular, given that for
our models we can compute the line-of-sight velocity distribution, we can use it 
to build up a Maximum Likelihood Estimator (MLE) based on measures of velocities 
of individual stars, thus eliminating any kind of information loss due to binning
the kinematic data (\citealt{Watkins2013}).

As a first application, in this paper we apply $f({\bf J})$ models to the
Fornax dSph, which was the first to be discovered (\citealt{Shapley1938}). Fornax 
is located at high Galactic latitude at a distance of $138\pm8\,$kpc
(\citealt{Mateo1998}; \citealt{Battaglia2006}), and has the largest body of
kinematic data. There are quantitative indications (\citealt{Battaglia2015}) that
the effect of the tidal field of the Milky Way on the present-day dynamics of Fornax
is negligible, so we are justified in modelling this galaxy as a stationary
isolated stellar system.

This paper is organised as follows: in Section \ref{sec:df} we introduce the
DF that we propose for dSphs and summarise the main characteristics of the models 
it generates. In Sections \ref{sec:Analysis2fj} models are compared to observations
of dSphs. In Section \ref{sec:fornax} we present the results obtained applying our
technique to the Fornax dSphs. Section \ref{sec:conc} concludes.

\section{two-component $f({\bf J})$ models for dwarf spheroidal galaxies}
\label{sec:df}

We model a dSph as a two-component system with stars and dark matter.  

\subsection{Stellar component}
\label{subsec:st}

The stellar component is described by the DF
\begin{equation}\label{for:df1}
\fJ = \frac{\Mfst}{\Jfst^3}\exp\biggl[-\biggl(\frac{\kJ}{\Jfst}\biggr)^{\alpha}\biggr],
\end{equation}
with 
\begin{equation}\label{for:df2}
\kJ = J_r + \eta_{\phi}|J_{\phi}| + \eta_zJ_z,
\end{equation}
where ${\bf J} = (J_r, J_{\phi}, J_z)$ comprises $J_r$, the radial action, $J_{\phi}$,
the azimuthal action, and $J_z$ the vertical action, $\Mfst$ is a characteristic 
mass, $\Jfst$ is a characteristic action, and $\alpha$, $\eta_{\phi}$ and $\eta_z$ 
are dimensionless, non-negative, parameters. The DF in equation (\ref{for:df1}) proves 
to be expedient in representing dSph since it generates an almost exponential cut-off 
in the density distribution, similar to what is observed for typical dSphs
(\citealt{IrwinHatzidimitriou1995}). 

\subsection{Dark-matter component}
\label{subsec:dm}

We consider a family of DFs for the dark halo such that, in the absence of baryons, the 
dark-matter density distribution is very similar to an exponentially truncated 
\citet*[hereafter NFW]{NavarroFrenkWhite1996} profile, with the optional presence of
a central core. Specifically, the dark-matter component is described by the DF
\begin{equation}\label{for:dmdf}
 \fJdm = \fJPst\gJ\tJ,
\end{equation}
where 
\begin{equation}\label{for:PostiDF}
 f({\bf J}) = \frac{\Mfdm}{\Jfdm^3}\frac{[1 + \Jfdm/\hJ]^{5/3}}{[1 + \hJ/\Jfdm]^{2.9}},
\end{equation}
\begin{equation}\label{for:ColeBinneyDF}
 \gJ = \biggl[\biggl(\frac{\Jcdm}{\hJ}\biggr)^2 -\mu\frac{\Jcdm}{\hJ} + 1\biggr]^{-5/6}
\end{equation}
and
\begin{equation}\label{for:expcoffDF}
 \tJ = \exp{\biggl[-\biggl(\frac{\hJ}{\Jtdm}\biggr)^{2}\biggr]}.
\end{equation}
Here, $\Mfdm$ is a characteristic mass scale and $\Jfdm$ is a characteristic action scale,
while $\hJ$ is the homogeneous function of the actions
\begin{equation}\label{for:dfPosti1}
 \hJ = J_r + \delta_{h, \phi}|J_{\phi}| + \delta_{h, z}J_z, 
\end{equation}
where $\delta_{h, \phi}$ and $\delta_{h,z}$ are dimensionless, non-negative, parameters 
regulating the velocity distribution of the halo. \cite{Posti2015} introduced the DF 
(\ref{for:PostiDF}) to describe NFW-like $f({\bf J})$ models \footnote{In \cite{Posti2015} two
different homogeneous functions are used in the numerator and in the denominator of the DF 
in order to have more freedom in the anisotropy profile of the model. Here we do not explore 
the anisotropy of the halo, so we can adopt a single homogeneous function $h$ as in equation 
(\ref{for:PostiDF}).}. To avoid the divergence of the dark-matter mass for large actions we 
multiply the DF by the exponential term (\ref{for:expcoffDF}), in which $\Jtdm$ is a 
characteristic action that determines the spatial truncation of the density distribution. 
Following \cite{ColeBinney2017}, in equation (\ref{for:dmdf}) the DF of \cite{Posti2015} is
multiplied by the function $\gJ$ in order to produce a core in the innermost regions of the
dark-matter density distribution. The size of the core is regulated by the characteristic
action $\Jcdm$. The dimensionless parameter $\mu$ is used to make the integral of the DF of 
(\ref{for:dmdf}) independent of $\Jcdm$: the value of $\mu$ is such that models with different 
$\Jcdm$, but with the same values of the other parameters of the DF (\ref{for:dmdf}), have 
the same total dark-matter mass. 

\subsection{General properties of the models}
\label{subsec:modpr}

The total mass of each component is fully determined by the properties of its DF and is 
independent of the presence and properties of the other component (\citealt{Binney2014}). 
The total stellar mass is

\begin{equation}\label{for:totmass}
 \Mstot = (2\pi)^3\int\fJ\text{d}^3{\bf J}, 
\end{equation}
while the total dark-matter mass is
\begin{equation}\label{for:totmassDM}
 \Mdmtot = (2\pi)^3\int\fJdm\text{d}^3{\bf J}.
\end{equation}
The stellar and dark-matter density distributions are, respectively,
\begin{equation}\label{for:rho1}
 \rhost({\bf x}) = \int \fJ {\rm d}^3 {\bf v} 
\end{equation}
and
\begin{equation}\label{for:rho2}
 \rhodm({\bf x}) = \int \fJdm {\rm d}^3 {\bf v}.
\end{equation}
Evaluation of the integrals (\ref{for:rho1}) and (\ref{for:rho2}) involves the evaluation of the 
action ${\bf J}$ as functions of the ordinary phase-space coordinates $({\bf x}, 
{\bf v})$ in the total gravitational potential $\Phitot = \Phist + \Phidm$, where $\Phist$ 
is the stellar gravitational potential, given by $\nabla^2 \Phist = 4\pi G\rhost$, and 
$\Phidm$ is the dark-matter gravitational potential, given by $\nabla^2\Phidm = 4\pi G\rhodm$. 
Thus, the problem is non-linear and the density-potential pairs $(\rhost, \Phist)$ and 
$(\rhodm, \Phidm)$ are computed iteratively (see \citealt{Binney2014} , \citealt{Posti2015}
and \citealt{Sanders2016}). Both DFs (\ref{for:df1}) and (\ref{for:dmdf}) are even in
$J_{\phi}$, so they define non rotating models. Putting any component in rotation is 
straightforward following, for instance, the procedure described in \cite{Binney2014}. 
For non-rotating models, the velocity dispersion tensor of the stellar component is
\begin{equation}\label{for:sigij}
\sigma^2_{i,j} \equiv \frac{\int v_iv_j\fJ {\rm d}^3{\bf v}}{\rhost(\bf x)},
\end{equation}
where $v_i$ and $v_j$ are the $i$-th and $j$-th components of the velocity.

The characteristic length and velocity scales of the stellar component are, respectively,
\begin{equation}\label{for:JdSMdS1}
 \rfst \equiv \frac{\Jfst^2}{G \Mfst}
\end{equation}
and
\begin{equation}\label{for:JdSMdS2}
 \vfst \equiv \frac{G\Mfst}{\Jfst}.
\end{equation}
The characteristic length and velocity scales of the dark halo are, respectively,
\begin{equation}\label{for:JdmMdm1}
 \rfdm \equiv \frac{\Jfdm^2}{G \Mfdm} = \biggl(\frac{\Jfdm}{\Jfst}\biggr)^2\frac{\Mfst}{\Mfdm}\rfst =
 \frac{\Jftil^2}{\Mftil}\rfst
\end{equation}
and
\begin{equation}\label{for:JdmMdm2}
 \vfdm \equiv \frac{G\Mfdm}{\Jfdm} = \frac{\Mfdm}{\Mfst}\frac{\Jfst}{\Jfdm}\vfst = \frac{\Mftil}
 {\Jftil}\vfst,
\end{equation}
where $\Mftil \equiv \Mfdm/\Mfst$ and $\Jftil \equiv \Jfdm/\Jfst$.

\subsection{Spherical models}
\label{sec:spm}

The simplest models belonging to the family described in Sections \ref{subsec:st} and 
\ref{subsec:dm} are those in which both the dark-matter and the stellar components are 
spherically symmetric ($\eta_\phi = \eta_z$ in equation \ref{for:df1}, and $\delta_{h, 
\phi} = \delta_{h,z}$, in equation \ref{for:dfPosti1}). In general neither component is
spherical if $\delta_{h,\phi}\neq\delta_{h,z}$ or $\eta_{\phi}\neq\eta_z$. Here we 
focus on the spherical case and define 
\begin{equation}
 \eta \equiv \eta_{\phi} = \eta_z
\end{equation}
and 
\begin{equation}
 \delta \equiv \delta_{\phi, h} = \delta_{z, h}. 
\end{equation}
We require the dark-matter velocity distribution to be almost isotropic setting 
$\delta \leq 1$ (\citealt{Posti2015}). With these assumptions each of our models depends on
the eight parameters  
\begin{equation}\label{for:param}
 \xib \equiv (\alpha, \eta, \Mftil, \Jftil, \Jctil, \Jttil, \Mfst, \Jfst), 
\end{equation}
where $\Jctil \equiv \Jcdm/\Jfdm$ and $\Jttil \equiv \Jtdm/\Jfdm$.
Models that share the dimensionless parameters $\alpha$, $\eta$, $\Mftil$, $\Jftil$,
$\Jctil$ and $\Jttil$ are homologous. The physical units are determined by the dimensional
parameters $\Mfst$ and $\Jfst$.

The stellar density distribution is characterised by an extended core and a truncation 
of adjustable steepness in the outskirts (see Section \ref{sec:fornax}). For the stellar 
component we define the half-mass radius $\rh$ as the radius of the sphere that contains 
half of the total stellar mass. The most general spherical $f({\bf J})$ model of Section 
\ref{subsec:dm} generates a dark-matter density profile characterised by three regimes: 
a core where the logarithmic slope of the density profile $\gamma \equiv 
\text{d}\ln\rho_{\rm dm}/\text{d}r \sim 0$, an intermediate region where $\gamma \sim -1$ 
and the outer region where $\gamma \sim -3$. For each model we define the core radius 
$\rc \equiv r_{-1/2}$ (radius at which $\gamma = -1/2$), the scale radius $\rs \equiv 
r_{-2}$ (radius at which $\gamma =-2$, as for the scale radius of the classical NFW model),
and the truncation radius $\rt \equiv r_{-3}$ (radius at which $\gamma = -3$).

The eight parameters $\xib$ (equation \ref{for:param}) are quantities appearing in the DFs 
(equations \ref{for:df1} and \ref{for:dmdf}) or combinations thereof (see Section 
\ref{subsec:modpr}). Once a model is computed, it can be also characterised by the eight 
parameters
\begin{equation}
 \xib' = (\alpha, \eta, \Mdmt, \rst, \rct, \rtt, \Mfst, \Jfst),
\end{equation}
where we have replaced $\Mftil$, $\Jftil$, $\Jctil$ and $\Jttil$ with $\Mdmt\equiv\Mdmtot/\Mstot$,
$\rst\equiv\rs/\rh$, $\rct\equiv\rc/\rh$, $\rtt\equiv\rt/\rh$, which have a more
straightforward physical interpretation. In the following we briefly comment on the six 
dimensionless parameters $\alpha$, $\eta$, $\Mdmt$, $\rst$, $\rct$ and $\rtt$.

\begin{itemize}
 \item $\alpha$: this mainly regulates the shape of the density profile of the stellar 
 component. We find empirically that for higher values of $\alpha$ the core is flatter 
 and the outer profile is steeper. This is expected because for higher values of $\alpha$
 the DF (\ref{for:df1}) is more rapidly truncated for large actions.
 
 \item $\eta$: this mainly regulates the velocity anisotropy of the stellar component. We 
 find empirically that higher values of $\eta$ generate more radially biased models. This 
 is expected because orbits with large $|J_{\phi}|$ or $J_z$ are penalized for large values 
 of $\eta$ (see equations \ref{for:df1} and \ref{for:df2}; we recall that for spherical models 
 $\eta_{\phi} = \eta_z = \eta$).
 
 \item $\Mdmt$: this is the ratio between the total dark-matter mass $\Mdmtot$ and the total
 mass of the stellar component $\Mstot$. Both $\Mdmtot$ and $\Mstot$ are well defined because 
 the integrals in equations (\ref{for:totmass}) and (\ref{for:totmassDM}) converge. Since the 
 DFs (\ref{for:df1}) and (\ref{for:dmdf}) depend on homogeneous functions of the actions, for 
 spherical models the total masses are given by the one-dimensional integrals (\citealt{Posti2015}) 
 
 \begin{equation}\label{for:dm1d}
  \frac{\Mdmtot}{\Mfdm} = \frac{(2\pi)^3}{\delta^2}\int_0^{\infty}h^2\fJdmh\text{d}h
 \end{equation}
 for the dark halo, and
 
 \begin{equation}\label{for:st1d}
  \frac{\Mstot}{\Mfst} = \frac{(2\pi)^3}{\eta^2}\int_0^{\infty}h^2\fJh\text{d}h
 \end{equation}
 for the stellar component (for details, see Appendix \ref{app:A}). Given that $\Mdmt = 
 \Mdmtot/\Mstot$, equations (\ref{for:dm1d}) and (\ref{for:st1d}) can be combined to give 
 
 \begin{equation}
  \Mdmt = \Mftil\frac{\eta^2}{\delta^2}\frac{\int_0^{\infty}h^2\fJdmh\text{d}h}{\int_0^{\infty}h^2\fJh\text{d}h}.
 \end{equation}
 The dark-matter to stellar mass ratio can be fixed by adjusting $\Mftil$ and $\mu$, the 
 normalization parameter appearing in the definition of $f_{\rm dm}$ (see equations \ref{for:dmdf}
 and \ref{for:ColeBinneyDF}).
 
 \item $\rst$: this is the ratio between the scale radius of the halo $\rs$ and the half-mass 
 radius of the stellar component $\rh$. For sufficiently large $\rst$, the dark-matter
 density profile is essentially a power law in the region populated by stars. This property
 makes the characteristic scale radius $\rs$ and the normalisation of the dark-matter 
 component degenerate: provided $\rst \gg 1$, dark-matter density profiles with different
 values of $\rs$ affect the stellar component in the same way, if properly scaled. Differently
 from $\Mdmt$, $\rst$ cannot be fixed a priori since it depends on the total gravitational
 potential $\Phitot$. However, a model with a predefined value of $\rst$ can be obtained 
 iteratively.
 
 \item $\rct$: this is the ratio between the core radius of the dark-matter component $\rc$
 and the half-mass radius of the stellar component $\rh$. $\rct$ cannot be fixed a priori
 because it depends on $\Phitot$. However, for the two-component models here considered,
 we find empirically that $\rct$ can anyway be fixed with reasonable precision by fixing 
 $\Jctil$.
 
 \item $\rtt$: this is the ratio between the truncation radius of the halo $\rt$ and the
 half-mass radius of the stellar component $\rh$. $\rtt$ depends on $\Phitot$, so it 
 cannot be fixed a priori. In general, models with the same value of truncation action
 $\Jttil$ do not have the same value of $\rtt$.
 
\end{itemize}










\section{STATISTICAL ANALYSIS}
\label{sec:Analysis2fj}
\subsection{Comparison with data}
\label{subsec:comparison}

When applying the spherical models presented in Section \ref{sec:spm} to an observed dSph
galaxy, the best model (i.e. the best set of eight parameters $\xib$) is determined through 
a comparison with a set of observables. The dSph may be elliptical on the sky while our
model will be  spherical, so we assign each star a circularised radius

\begin{equation}\label{eq:circularise}
 R \equiv \sqrt{x^2(1-\epsilon) + \frac{y^2}{(1-\epsilon)}},
\end{equation}
where $\epsilon\equiv 1 - b/a$, with $b$ and $a$  the lengths of the semi-minor and semi-major 
axes, is the ellipticity of the galaxy's image on the sky and $(x,y)$ are the star's
Cartesian coordinates in the reference frame aligned with the image's principal axes. 

We assume the data comprises a photometric sample, used to compute the projected stellar 
number density $\nobs$, and a kinematic sample with measurements of the line-of-sight 
velocities $\vlos$ of individual stars.  We refer to the observed number density as
a set of $N_n$ observed values $\{R_i, \nobsi\}$, with $i=1,..,N_n$, and to the
line-of-sight velocities as $N_v$ measures $\{R_k, \vlosk \}$, with $k=1,..,N_v$.
For each model we compute the stellar surface number density distribution

\begin{equation}\label{for:Iproj}
 \nmod({\bf x}_{\perp}) = \frac{N_{\rm tot, \star}}{\Mstot}\int\rhost({\bf x})\text{d}x_{||},
\end{equation}
where $N_{\rm tot, \star}$ is the total number of stars of the photometric sample, 
and the model line-of-sight velocity distribution (hereafter LOSVD)

\begin{equation}\label{for:LOSVD}
 {\mathscr L}_{\star}({\bf x}_{\perp}, v_{||}) = \frac{\int f_{\star}[{\bf J}({\bf x},{\bf v})]\text{d}x_{||}\text{d}{\bf v}_{\perp}}
 {\rho_{\star}({\bf x}_{\perp})}.
\end{equation}
Here, $x_{||} \equiv {\bf x} \cdot {\bf \hat{s}}$ and ${\bf x}_{\perp} = {\bf x} - 
x_{||}{\bf \hat{s}}$ are, respectively, the parallel and orthogonal components of the
position vector with respect to the line-of-sight (unit) vector ${\bf \hat{s}}$, and
$v_{||}$ is the velocity component along ${\bf \hat{s}}$. For spherical models $\nmod$
and ${\mathscr L}_{\star}$ depend on ${\bf x}_{\perp}$ only through the scalar projected
distance from the center on the plane of sky $R \equiv ||{\bf x}_{\perp}||$. 

We compare models to data with a maximum likelihood method. The log-likelihood of a
model is defined as 

\begin{equation}\label{for:logl}
 \ln \Ltot = \ln \Ln + \ln \Lv,
\end{equation}
with

\begin{equation}
 \ln \Ln = -\frac{1}{2}\sum_{i=1}^{N_{n}}\biggl(\frac{\nobsi - \nmod(R_i)}{\delta n_i}\biggr)^2,
\end{equation}
where $\delta n_i$ are the uncertainties of the stellar number density measurements, and

\begin{equation}\label{for:lLv}
 \ln \Lv = \sum_{k=1}^{N_v}\ln(p_{v,k}).
\end{equation}
In the above equation

\begin{equation}
 p_{v,k} \equiv \int_{-\infty}^{+\infty}{\mathscr L}_{\rm tot}(R_k, v_{||}) G_k(v_{||} - \vlosk)\text{d}v_{||} 
\end{equation}
is the convolution of the total LOSVD ${\mathscr L}_{\rm tot}$ and a Gaussian distribution
$G_k$ with null mean and standard deviation equal to the uncertainty on the line-of-sight
velocity of the $k$-th star. The total LOSVD

\begin{equation}\label{for:Lv}
 {\mathscr L}_{\rm tot} \equiv (1-\omega_k){\mathscr L}_{\star} + \omega_k{\mathscr L}_{{\rm f}, k}
\end{equation}
accounts for the fact that the kinematic sample of stars may be contaminated by field stars:

\begin{equation}
{\mathscr L}_{{\rm f}, k} \equiv {\mathscr L}_{{\rm f}} (\vlosk)
\end{equation}
is the LOSVD ${\mathscr L}_{\rm f}$ of field stars evaluated at $\vlosk$ and

\begin{equation}\label{for:omegak}
 \omega_k \equiv \frac{n_{\rm f}}{\nobs(R_k) + n_{\rm f}}
\end{equation}
weights the relative contribution between dSph and contaminants. $n_{\rm f}$ is the
mean projected number density of field stars, which is taken to be constant throughout 
the extent of the galaxy, while $\nobs(R_k)$ is the observed projected number density 
profile evaluated at $R_k$. 

\subsection{Models and families of models}
\label{subsec:models}

\begin{table}
 \begin{center}
  \begin{tabular}{ccccccc}
  \hline\hline
  $\Delta\ln\Ltot_{j,m}$	&	$j=1$	&	$j=2$	&	$j=3$	&	$j=4$	&	$j=5$	&	$j=6$	\\
  \hline\hline
      $m=1$			&	0.50	&	1.15	&	1.77	&	2.36	&	2.95	&	3.52	\\
      $m=2$			&	2.00	&	4.01	&	4.85	&	4.85	&	5.65	&	6.40	\\
      $m=3$			&	3.00	&	5.90	&	7.10	&	8.15	&	9.10	&	10.05	\\
  \hline\hline
  \end{tabular}
  \caption{Values of the delta log-likelihood $\Delta\ln\Ltot_{j,m}$ (equation \ref{for:confcont})
  corresponding to $m\sigma$ confidence levels. $j$ is the number of free parameters of a family of
  models.}\label{tab:chi2}
 \end{center}
\end{table}

In the terminology used in this work, we distinguish the terms {\em model} and {\em 
family of models}. We refer to a class of spherical systems with the same values of the 
six dimensionless parameters ($\alpha$, $\eta$, $\Mftil$, $\Jftil$, $\Jctil$, $\Jttil$)
as a model. Each model maps a two dimensional sub-space ($\Jfst, \Mfst$) of homogeneous 
systems. When a model is compared with observations, we find the values of $\Jfst$ and 
$\Mfst$ that maximise $\mathcal{L}$ (equation \ref{for:logl}) and, with a slight abuse 
of the terminology, we define its likelihood
as this maximum value of $\mathcal{L}$. 

We will refer to a set of models sharing some properties (i.e. values of some parameters)
as a family of models. For instance, we will define the family of one-component (or
mass-follows-light, MFL) models as the set of all models with $\Mfdm = 0$. Each family
of models has $j$ free parameters, which we indicate with the $j$-dimensional vector 
$\xib_j$. For instance, for spherical MFL models $j=4$ and $\xib_4$ = ($\alpha$, $\eta$, 
$\Jfst$, $\Mfst$). The best model of a family is the model with the maximum likelihood 
among all those belonging to that family. 



 
For each family we explore the parameter space using as stochastic search 
method a Markov-Chain Monte Carlo (MCMC) algorithm based on a Metropolis-Hastings sampler 
(\citealt{Metropolis1953}, \citealt{Hastings1970}) to sample from the posterior distribution
using uninformative priors on the parameters. In each case we find that the MCMC 
allows us to finely sample the relevant region of the parameter space, including 
the best model and all the models within $1\sigma$.
For a given family, the $m\sigma$ confidence levels ($m=1,2,3...$) on any quantity 
(and thus the uncertainty bands in the plots)  are constructed by selecting in the 
parameter space $\xib_j$ all models with likelihood such that

\begin{equation}\label{for:confcont}
\ln \Ltot_{\rm max} - \ln \Ltot(\xib_j) < \Delta\ln\Ltot_{j,m},
\end{equation}
where $\ln \Ltot_{\rm max}$ is the log-likelihood of the best model of the family and 
$\Delta\ln\Ltot_{j,m}$ is a threshold value of $\Delta\ln \mathcal{L}$ depending on $j$ and 
$m$. Reference values of $\Delta \ln \Ltot_{j,m}$, relevant for the cases
considered in this work, are given in Table \ref{tab:chi2}.

To estimate the relative goodness of different families of models, with possibly different
numbers of free parameters, we use the Akaike Information Criterion (AIC, \citealt{Akaike1973}). 
Given $\Ltot_{\rm max}$, the maximum likelihood of a family with $j$ free parameters, we 
define the quantity

\begin{equation}\label{for:AIC}
 {\rm AIC} = 2j - 2\ln \Ltot_{\rm max}
\end{equation}
as a measure of the goodness of the best model of the family, which takes into account 
the number of free parameters.  Among all families, the best model is the one with the
minimum value of AIC $({\rm AIC}_{\rm best})$ and

\begin{equation}\label{for:prob}
 P \equiv \exp[({\rm AIC}_{\rm best} - {\rm AIC})/2]
=\exp(j_{\rm best}-j)\frac{\Ltot_{\rm max}}{\Ltot_{\rm max,\rm best}}
\end{equation}
is the probability that the best model of another family represents the data as well as the 
best model of all models (here, $j_{\rm best}$ and $\Ltot_{\rm max,\rm best}$ are, respectively, 
the number of free parameters and the likelihood of the best of all models).

\section{APPLICATION TO FORNAX}
\label{sec:fornax}

\subsection{Data set}
\label{subsec:dataset}


\begin{figure*}
 \begin{center} 
  \includegraphics[width=1.\hsize]{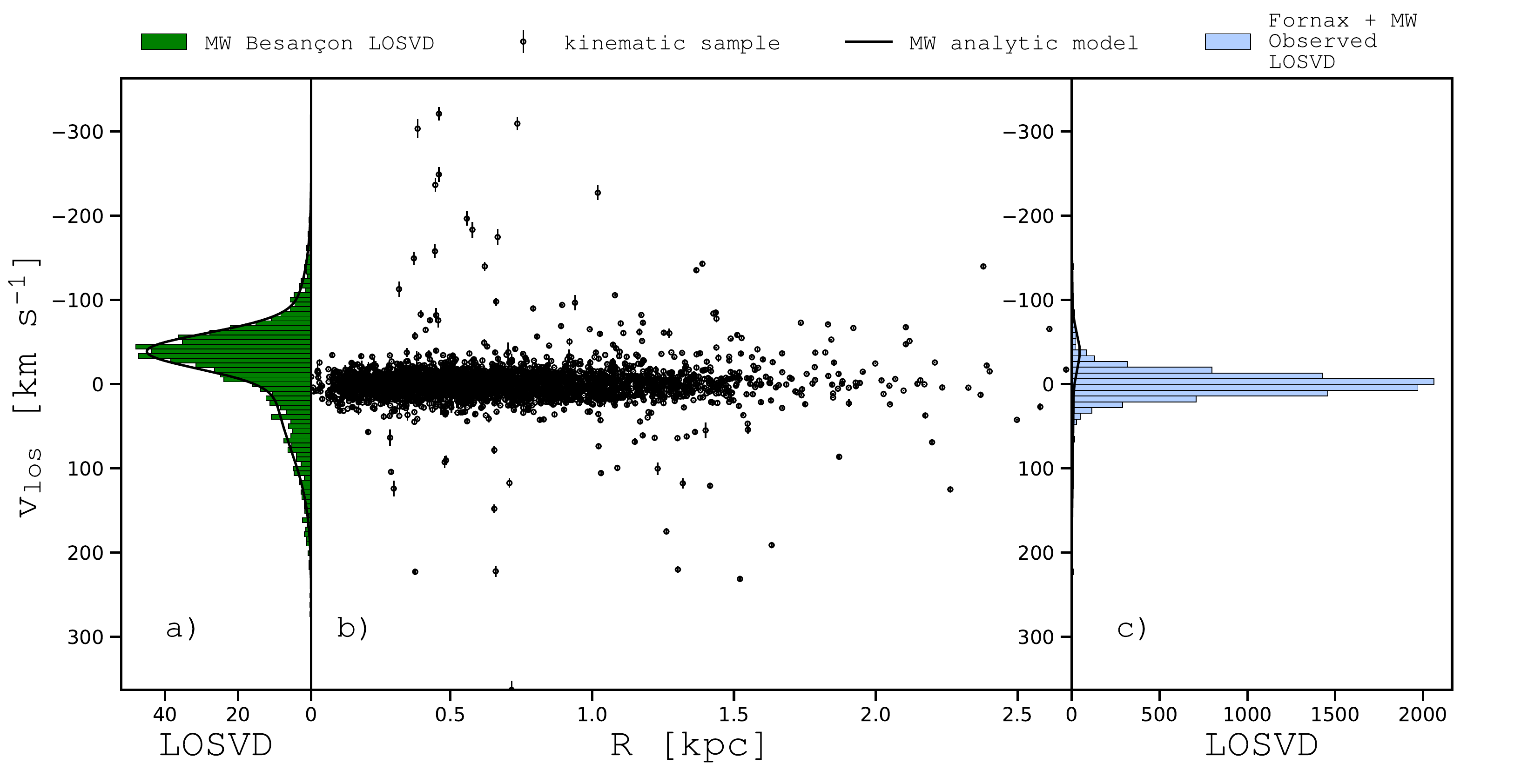}
  \caption{Panel a: The Milky Way's LOSVD in the direction of Fornax calculated from the
  Besan\c{c}on model (histogram) and the best-fitting two-Gaussian distribution (black line).
  The LOSVD of the Besan\c{c}on model has been shifted by $v_{\rm sys}$ to make the comparison 
  coherent. Panel b: position-velocity diagram of the whole kinematic sample used in this work 
  (Fornax + MW; data taken from \citealt{Battaglia2006} and \citealt{Walker2009}; see text). 
  Panel c: Fornax + MW observed LOSVD superimposed to the best fitting two-Gaussian distribution
  of panel a. The observed total LOSVD (Fornax + MW) of panel c is normalized to the total number 
  of stars (Fornax + MW) expected from the \citet{Battaglia2006} photometric sample, while the 
  MW model of panels a and c is normalized to the total number of the field stars expected in the
  very same region according to the Besan\c{c}on model. Note the very different scales of the 
  x-axis of panels a and c.}\label{fig:posvel}
 \end{center}
\end{figure*}

\begin{table}
\begin{center}
\begin{tabular}{lcc}
 \hline\hline
 Parameter				&	Value					& Reference 	\\
 \hline\hline
 RA					&	2$^h$ 39$^m$ 52$^s$			&	1	\\
 DEC 					&	-34\textdegree 30' 49''			&	1	\\
 P.A.					&	46.8\textdegree$\pm$1.6\textdegree	&	1	\\
 $\epsilon$				&	0.30$\pm$0.01				&	1	\\
 $d$	[kpc]				&	138					&	1	\\
 $\Reff$  [kpc]				&	0.62					&	1	\\
 $N_n$					&	27					&	1	\\				
 $n_{\rm f}$ [stars arcmin$^{-2}$]	&	0.263					&	1	\\
 $v_{\rm sys}\quad[\kms]$		&	55.1					&	2	\\
 $N_v$					&	2990					&	3	\\
 \hline\hline
\end{tabular}
\caption{Values of the main observational parameters of Fornax used in this work: right 
ascension (RA), declination (DEC), Position Angle (P.A.), ellipticity ($\epsilon$), distance 
from the sun ($d$), projected half-light radius ($\Reff$), number of bins of the projected 
stellar number density profile ($N_n$), mean projected number density of the field stars 
($n_{\rm f}$), systemic heliocentric velocity ($v_{\rm sys}$), number of members of the 
kinematic sample ($N_v$). References: 1) \citealt{Battaglia2006}, 2) \citealt{BreddelsHelmi2013}, 
3) this work.}\label{tab:FnxParam}
\end{center}
\end{table}

Our photometric sample is taken from \cite{Battaglia2006}, who, using deep ESO/WIFI observations, 
studied the spatial distribution of the stars of Fornax and derived its main structural 
parameters.  Adopting a distance $d=138$ kpc (\citealt{Battaglia2006}), the projected 
stellar number density profile extends out to $3.33\,$kpc and it is composed of $N_n=27$ 
concentric elliptical shells of semi-major axis length $ R_{i, {\rm ell}}$ of equal
thickness, so $R_{i+1,{\rm ell}}-R_{i,{\rm ell}}= 0.12\,$kpc for all $i$. The shells
have ellipticity $\epsilon=0.3$ (\citealt{Battaglia2006}).  We use the observed
projected stellar number density profile as a function of the circularized radius 
$R_i \equiv R_{i, {\rm ell}}\sqrt{1-\epsilon}$ with $i=1,..,N_n$. The circularized 
projected half-light radius is $\Reff = 0.62\,$kpc. 

\begin{table*}
\begin{center}
\renewcommand{\arraystretch}{1.8}
\begin{tabular}{p{1.7cm}cccccccc}
\hline\hline
 Family		& $\alpha$	& $\eta$	&	$\Mftil$	&	$\Jftil$ 	&	$\Jctil$ 	&	$\Jttil$	&	$\Jfst$/[km s$^{-1}$ kpc]	&	$\Mfst$/[$M_{\odot}$]	\\
 \hline\hline
 FnxMFL		& $1.52^{+0.03}_{-0.04}$ & $0.49^{+0.02}_{-0.03}$ & 0 &  --  &  --     &  --  & $6.87^{+0.28}_{-0.44}$ & $7.70^{+0.76}_{-1.09}\times10^7$ \\	
 
 FnxNFW		& $1.39^{+0.02}_{-0.03}$ & $0.38^{+0.02}_{-0.02}$ & $2.26^{+0.44}_{-0.41}\times10^2$ & $76.49^{+4.21}_{-3.85}$ &  --  &  6 & $5.00^{+0.35}_{-0.28}$ & $1.70^{+0.37}_{-0.27}\times10^7$ \\ 
 
 FnxCore1	& $0.84^{+0.02}_{-0.02}$ & $0.49^{+0.03}_{-0.03}$ & $1.56^{+0.28}_{-0.39}\times10^4$ & $196.58^{+15.43}_{-21.02}$ &  0.02     &  6 & $2.19^{+0.27}_{-0.16}$ & $5.52^{+1.81}_{-0.87}\times10^5$ \\ 
 
 FnxCore2	& $0.65^{+0.02}_{-0.02}$ & $0.56^{+0.04}_{-0.03}$ & $6.23^{+2.11}_{-2.14}\times10^4$ & $290.18^{+39.68}_{-40.34}$ &  0.20     &  6 & $0.98^{+0.16}_{-0.12}$ & $1.46^{+0.73}_{-0.38}\times10^5$ \\ 
 
 FnxCore3 (Best Model)	& $0.62^{+0.02}_{-0.01}$ & $0.56^{+0.04}_{-0.02}$ & $5.87^{+0.93}_{-2.22}\times10^4$ & $177.08^{+15.80}_{-29.59}$ &  0.67    & 6  & $0.84^{+0.17}_{-0.07}$ & $1.06^{+0.68}_{-0.10}\times10^5$ \\
 \hline\hline
\end{tabular}
\caption{Input parameters of the best Fornax models of each family. $\alpha$ and $\eta$: 
parameters of the stellar DF (\ref{for:df1}). $\Mftil \equiv \Mfdm/\Mfst$. $\Jftil\equiv\Jfdm/\Jfst$. 
$\Jctil\equiv\Jcdm/\Jfdm$. $\Jttil\equiv\Jtdm/\Jfdm$. $\Jfst$ and $\Mfst$: respectively, 
action and mass scales (equation \ref{for:df1}). $\Mfdm$, $\Jfdm$, $\Jcdm$ and $\Jtdm$ are 
the parameters of the dark-matter DF (equations \ref{for:dmdf}-\ref{for:expcoffDF}). The best
model is the FnxCore3.}\label{tab:bestParam}
\end{center}
\end{table*}

\begin{table*}
 \begin{center}
 \renewcommand{\arraystretch}{1.8}
 \begin{tabular}{p{1.7cm}ccccccccc}
  \hline\hline
  Family	& $\rct$	 &    $\Mdmt$ &  $\Mstot$ $/[M_{\odot}]$  &	$\Mrat$ & $\beta|_{1 {\rm kpc}}$& $\ln \Ltot_{\rm max}$ & ${\rm AIC}$ & $\Delta {\rm AIC}$ & $P$\\
  \hline\hline
  FnxMFL   	&        --      &     --        & $2.06^{+0.13}_{-0.12}\times10^8$  &   -- 	& $-0.32^{+0.13}_{-0.16}$  &  -12605.88  &  25219.76  &  185.74  &  $4.65\times10^{-41}$  \\
 
  FnxNFW	&        --      &  $63^{+14}_{-6}$ & $9.23^{+0.77}_{-2.85} \times 10^7$  &  $2.6^{+2.3}_{-0.8}$ &  $-0.73^{+0.23}_{-0.29}$ & -12582.16 & 25174.32 & 140.3 & $3.4\times10^{-31}$ \\	
 
  FnxCore1	&  $0.425^{+0.001}_{-0.012}$ & $1301^{+7}_{-164}$ & $1.00^{+0.73}_{-0.00}\times10^7$ & $73^{+1}_{-33}$ & $-0.17^{+0.15}_{-0.14}$ &  -12530.26  &  25070.52  &  36.5  &  $1.2\times10^{-8}$  \\
 
  FnxCore2	&  $1.075^{+0.001}_{-0.053}$ & $1344^{+38}_{-280}$ & $1.03^{+1.37}_{-0.03}\times10^7$ & $125^{+5}_{-76}$ & $0.07^{+0.12}_{-0.13}$ &  -12512.66  & 25035.32  &  1.3  &  0.52  \\
 
  FnxCore3  (Best Model) & $1.272^{-0.001}_{-0.035}$ & $946^{+1}_{-213}$ & $1.00^{+1.34}_{-0.00}\times10^7$  & $144^{+2}_{-87}$ & $0.08^{+0.14}_{-0.12}$ &  -12512.01  &  25034.02  &  0  &  1 \\ 
  \hline\hline
  \end{tabular}
  \caption{Output parameters of the best Fornax models of each family. $\rct$: ratio between 
  the core radius of the dark matter and the half-mass radius of the stellar component. $\Mdmt
  \equiv \Mdmt/\Mstot$. $\Mstot$: stellar total mass. $\Mrat$: dark-matter to stellar mass ratio
  within 3 kpc. $\beta|_{1 {\rm kpc}}$: anisotropy parameter (equation \ref{for:beta}) measured 
  at 1 kpc. $\ln \Ltot_{\rm max}$: log-likelihood (equation \ref{for:logl}). ${\rm AIC}$: value of the Akaike
  Information Criterion (equation \ref{for:AIC}). $\Delta$AIC: difference between the AIC of the 
  best model of a family and the best of all models (FnxCore3). $P$: probability that a model represents the data
  as well as the best in any family (FnxCore3). All models have $\rst = 4$ and 
  $\rh \simeq 0.81\,$kpc.}\label{tab:bestoutp}
 \end{center}
\end{table*}

Our reference kinematic sample of Fornax's stars is taken from \cite{Battaglia2006} and 
\cite{Walker2009}. This joined sample has already been used by \cite{BreddelsHelmi2013},
who corrected the line-of-sight velocities for the systemic velocity of Fornax
$v_{\rm sys}$ and for the gradient due to the extent of Fornax on the sky (for details 
see Table \ref{tab:FnxParam} and \citealt{BreddelsHelmi2013}). We apply the same corrections
here. The samples have been cross-matched with an astrometric precision of 1 arcsec and, 
for each duplicate (i.e. stars with two measured velocities), being $\delta v_1$ and 
$\delta v_2$ the different velocity errors of the cross-matched stars, we compute the 
average error

\begin{equation}
 \delta v = \sqrt{\frac{\delta v_1^2 + \delta v_2^2}{2}}.
\end{equation}
If the difference between the two velocities is larger than 3$\delta v$, we exclude the 
star from both the samples since we consider the difference to be caused by an unresolved 
binary. Otherwise, we use the mean of the two velocities. From the 945 stars of the 
\cite{Battaglia2006} sample and the 2633 of the \cite{Walker2009} sample, we find 488
cross-matched stars, 100 of which ($\approx$20\%) we classify binaries and thus exclude. 
In this way, the final kinematic sample consists of 2990 stars, each of which characterised by its 
line-of-sight velocity $\vlosk$ and its circularised radius $R_k$ (equation \ref{eq:circularise}).

Of course, our kinematic sample is still contaminated by undetected binaries. 
For instance, we expect to have in our sample about 600 undetected binaries ($\approx$20\%
of the non-cross matched stars) with properties similar to those excluded from the cross-matched
sample. Therefore we must quantify the effect of binary contamination 
on the LOSVD of our spectroscopic sample of Fornax. The contamination from undetected binaries is 
problematic when the characteristic velocity of short-period binaries is comparable with the 
line-of-sight velocity dispersion. \cite{Minor2010} found that for dwarfs with 
mean line-of-sight velocity dispersion in the range $4 \lesssim \sigma_{\rm los}$/km s$^{-1}$ 
$\lesssim 10$ the velocity dispersion profile may be inflated by no more than 15\% by undetected 
binaries, so binaries should have a negligible effect on Fornax, which has $\sigma_{\rm los} 
\simeq 12$ km s$^{-1}$.

In principle, though negligibly affecting $\sigma_{\rm los}$, the binaries could have 
an impact on the observed LOSVD. We tried to quantify this effect as follows: we built two kinematic 
samples, one containing all the cross-matched stars (488 stars; sample A) and one containing only 
stars not classified as binaries according to the above criterion
(388 stars; sample B). For these two samples we computed the LOSVD in two radial bins
($R<0.72$ kpc and $R>0.72$ kpc), such that each bin contains 244 stars in the case of sample A. 
According to the Kolmogorov-Smirnov test, in both radial bins the probability that the LOSVDs of samples 
A and B differ is less than 4\%. This result indicates that the LOSVDs used in our 
analysis should not be biased by the presence of undetected binaries.


The fields of view in the direction of Fornax suffer from significant Galactic contamination:
the mean velocity of MW stars in these fields is approximately the same as the systemic
velocity of Fornax, which complicates the selection of a reliable sample of members. 
From Fig.~\ref{fig:posvel}b, showing the position-velocity diagram of our kinematic 
sample, and from Fig.~\ref{fig:posvel}a, showing the velocity distribution of the MW 
calculated from the Besan\c{c}on model (\citealt{Besanson2004}) with a selection in
magnitude comparable to the one of our kinematic sample ($18\lesssim V \lesssim20.5$, 
with $V$ apparent $V$ band magnitude), we see that the LOSVDs of Fornax and MW stars
overlap (see also Fig.~\ref{fig:posvel}c).

As explained in Section \ref{subsec:comparison}, we take into account contamination by 
the MW by adding to our models a component describing the LOSVD of MW stars in the 
direction of Fornax. The MW velocity distribution extracted from the Besan\c{c}on
model is fitted with a two-Gaussian distribution (Fig.~\ref{fig:posvel}a) which reflects
the separate contributions of disc and halo stars. We assume a mean MW surface density 
$n_{\rm f} = 0.263$ stars arcmin$^{-2}$, obtained from the Besan\c{c}on model,
applying the the same selection in the $V$-band apparent magnitude as in the kinematic sample 
($18\lesssim V \lesssim20.5$). A summary of the main observational parameters of Fornax 
used in this work is given in Table \ref{tab:FnxParam}.

\subsection{Results}
\label{sec:res}

\begin{figure*}
 \centering 
 \begin{multicols}{2}
 \includegraphics[width=1.\hsize]{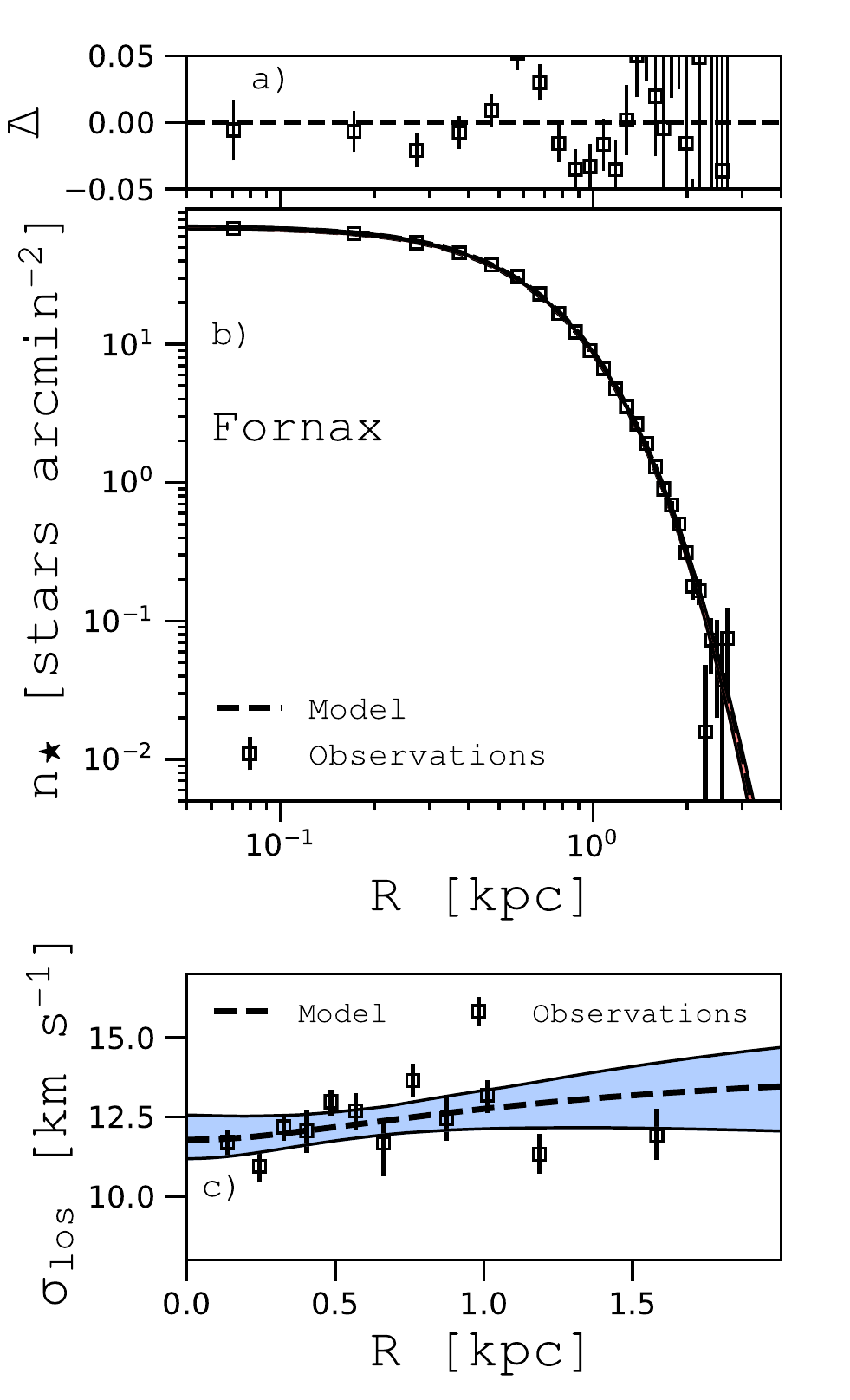} 
 \caption{Panel a: residuals $\Delta = (\nobs - n_{\star})/n_{\star}$ between the best model
 (FnxCore3) and the observed projected stellar number density profiles (dashed curve).
 Panel b: projected number density profile of the best model (dashed line) compared with 
 the observed profile (points with error bars). Panel c: line-of-sight velocity dispersion 
 profile of the best model compared with the observed profile (points with error bars). 
 Bands show the 1$\sigma$ uncertainty (see Section \ref{subsec:models}). Note that
 the x-axis is logarithmic in panel b and linear in panel c.}\label{fig:best_obs}
 \includegraphics[width=1.\hsize]{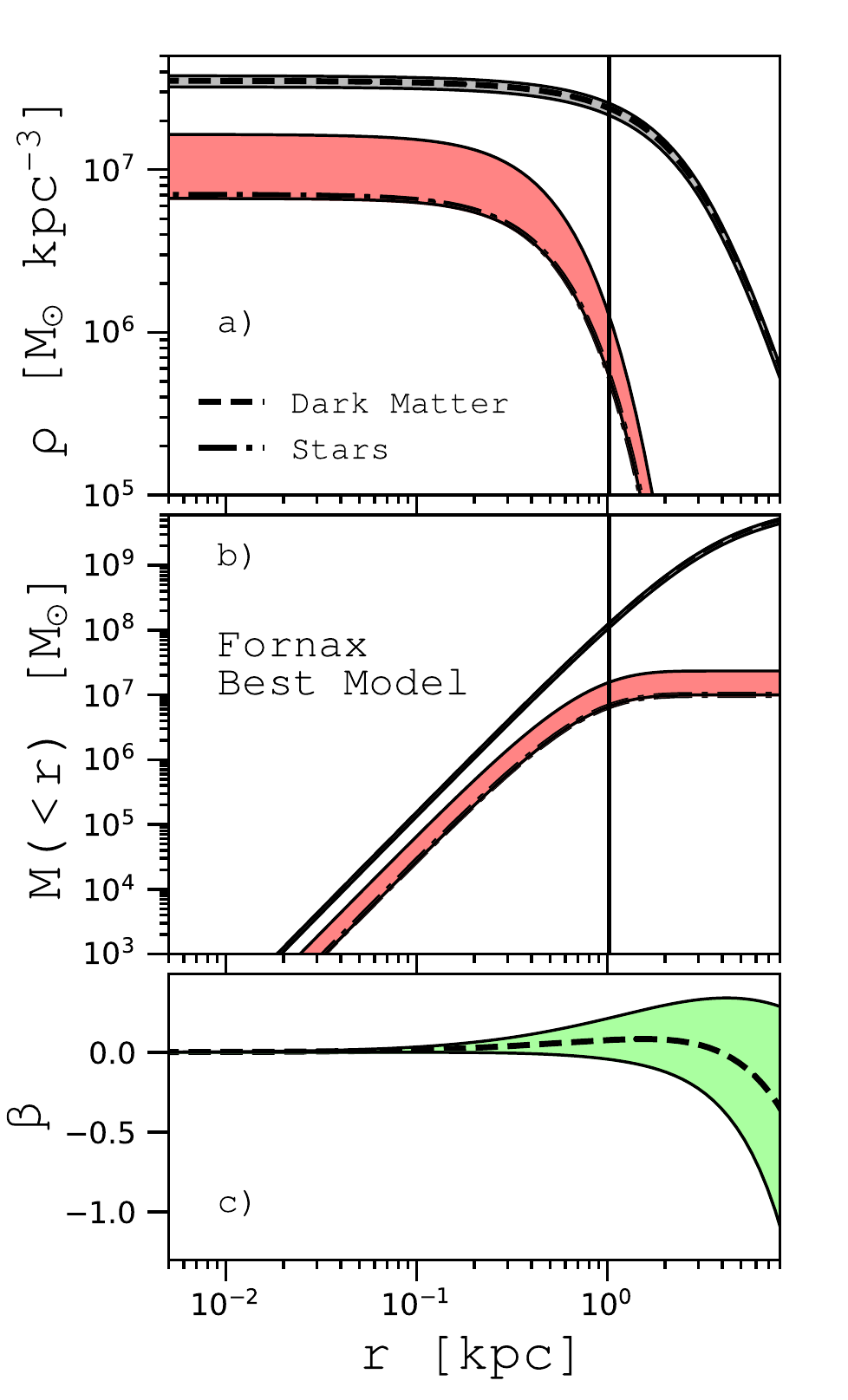}
 \caption{Panel a: stellar (dash-dotted line) and dark-matter (dashed line) density profiles
 of the best model (FnxCore3) of Fornax. Panel b: stellar (dash-dotted line) and dark-matter 
 (dashed line) mass profiles of the best model of Fornax. Panel c: stellar anisotropy parameter profile
 (dashed line) of the best model of Fornax. In panels a and b, the vertical lines mark the range 
 of the halo core radius $\rs$. The bands indicate the 1$\sigma$ uncertainty (see Section
 \ref{subsec:models}).}\label{fig:best_out} 
 \end{multicols}
\end{figure*}

Here we present the results we obtained applying the $f(\bf J)$ models of Section 
\ref{sec:df} to the Fornax dSph. In particular we focus on two-component spherical
models, in which the stars and the dark matter have different DFs. In Section 
\ref{subsec:onec} we will consider also simpler one-component spherical models, in 
which mass follows light. The physical properties of the models are computed by 
integrating equations (\ref{for:rho1}), (\ref{for:rho2}), (\ref{for:sigij}), 
(\ref{for:Iproj}) and (\ref{for:LOSVD}), using a code based on AGAMA (Action-based 
Galaxy Models Architecture, https://github.com/GalacticDynamics-Oxford/Agama; 
\citealt{Vasiliev2018}), a software package that implements the action/angle formalism of
$f({\bf J})$ DFs. To test the performances of our method, in the Appendix \ref{app:mock}
we applied $f({\bf J})$ models to a mock galaxy with structural and kinematic properties 
similar to a typical dSph.

\begin{figure*}
 \centering 
 \includegraphics[width=1.\hsize]{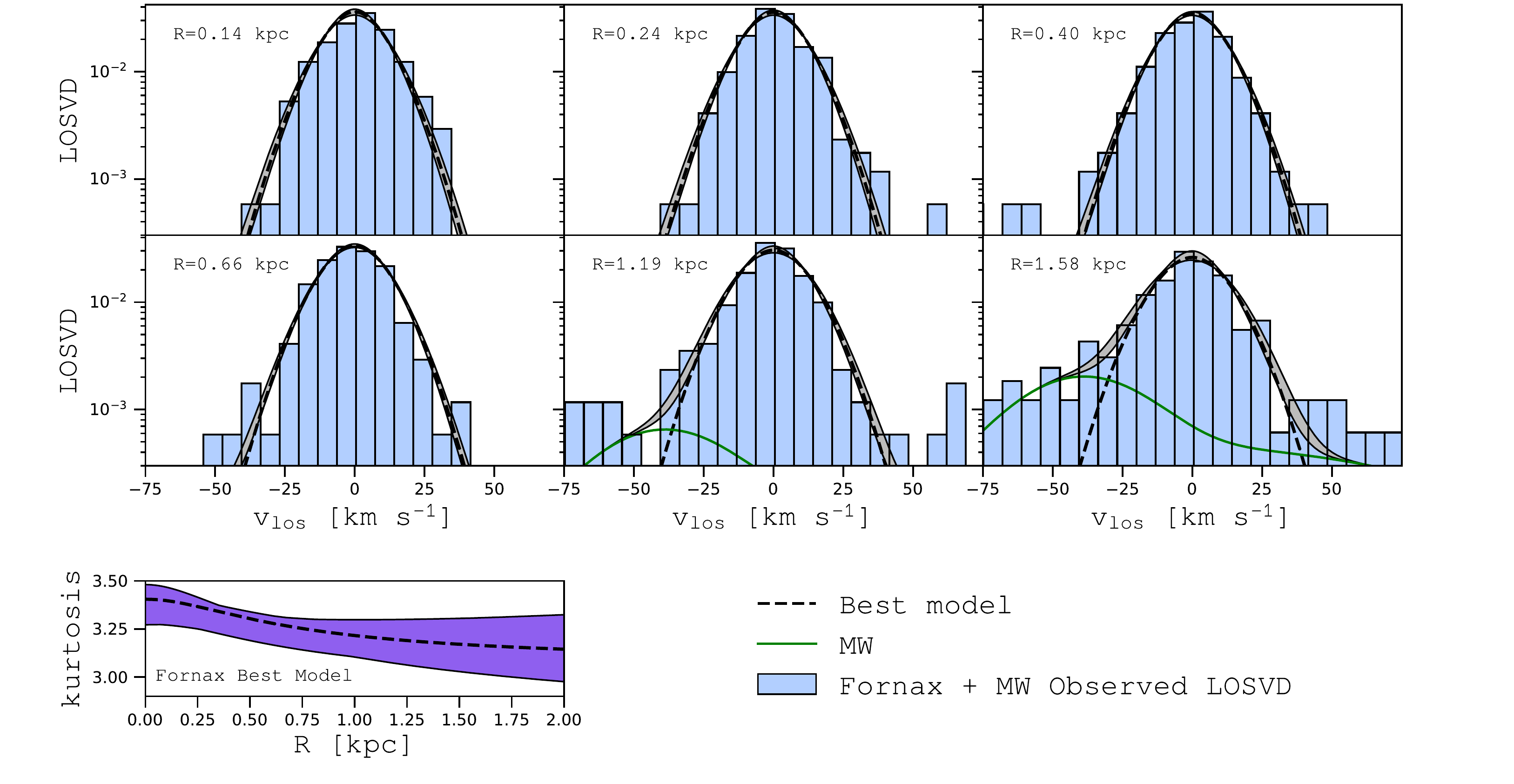} 
 \caption{Observed Fornax + MW LOSVD (histograms) compared with the  LOSVD of the best 
 model (FnxCore3). In each panel, $R$ indicates the average radius of the radial bin 
 where the LOSVD of the best model is computed. The radial bins are those used to construct 
 the observed line-of-sight velocity dispersion profile of Fornax (Fig.~\ref{fig:best_obs}). 
 The green solid curve marks the MW's contribution. The bottom panel shows the kurtosis
 profile of the best model's LOSVD. The bands mark the 1$\sigma$ uncertainty (see Section
 \ref{subsec:models}).}\label{fig:best_vel}
\end{figure*}

In the two-component models of Fornax, we adopt four families of dark halos: a family 
with a cuspy NFW-like halo and three halo families with central cores. Outside the
core region these fall off similarly to an NFW profile. For clarity, in the following 
we will refer to the cuspy NFW family as FnxNFW, and to the cored families as 
FnxCore$n$, with $n = 1,2,3$, where higher $n$ indicate larger cores in the dark halo.
The NFW halo is obtained setting $\Jctil = 0$ in equation (\ref{for:ColeBinneyDF}), 
while increasing values of $\Jctil$ produce cores of increasing sizes. The families 
FnxCore1, FnxCore2 and FnxCore3 have, respectively $\rct \simeq 0.43, 1.08, 1.28$,
corresponding to physical core radii $\rc \simeq 0.34, 0.87, 1.03$ kpc (see Section 
\ref{sec:spm}). We recall that the circularised projected half-light radius of Fornax
is $\Reff = 0.62$ kpc (Section \ref{sec:fornax}, Table \ref{tab:FnxParam}).
Based on observational estimates of the total stellar mass of Fornax (\citealt{deBoer2012}), 
we consider only two-component models such that $10^7\leq \Mstot/M_{\odot}\leq10^8$. 
We recall that the model $\Mstot$ depends only on $\alpha$, $\eta$ and $\Mfst$. 
Therefore, the above limits on $\Mstot$ are in practice limits on $\Mfst$, for given
$\alpha$ and $\eta$. We fixed
the ratio between the scale radius of the dark halo and the half-mass radius of the
stellar component to $\rst = 4$, consistent with the values expected on the basis 
of the stellar-to-halo mass relation and the halo mass-concentration relation, for 
galaxies with stellar masses $10^7\leq \Mstot/M_{\odot}\leq10^8$ (see Section 
\ref{subsec:insens}). We find that spherical models of Fornax have intrinsic stellar 
half-mass radius $\rh\simeq0.81$ kpc. 
It follows that our models have $\rs = \rst\rh \simeq 3.3$ kpc. Under these assumptions, 
each family has 5 free parameters, ($\alpha$, $\eta$, $\Mdmt$, $\Jfst$, $\Mfst)$.
Tables \ref{tab:bestParam} lists the values of the five parameters for 
the best model of each family, together with $\Jftil$ (fixed by the condition $\rst = 4$),
$\Jctil$ (fixed for each family) and $\Jttil=6$ for all families. The choice of 
$\Jttil=6$ ensures, for all the families, that the truncation radius of the dark halo 
$\rtt$ is much larger than the scale radius $\rst$. Table \ref{tab:bestoutp} 
gives some output parameters of the best Fornax model of each family.

\subsubsection{Properties of the best model}
\label{subsec:twocb}

According to our MLE (Section \ref{sec:Analysis2fj}), the best model belongs to the 
FnxCore3 family, with the most extended core in the dark-matter density profile ($\rc 
\simeq 1.03\,$kpc). In general, we find that a model in any cored families is strongly 
preferred to a NFW halo: the AICs (see Table \ref{tab:bestoutp}) indicate that the
introduction of even a small core in the dark-matter profile vastly improves the fit 
to the Fornax data.

Fig.~\ref{fig:best_obs}b plots the projected stellar number density profile of the best
model compared to the observed profile. The residuals between data and model are shown 
in Fig.~\ref{fig:best_obs}a. Fig.~\ref{fig:best_obs}c shows the line-of-sight velocity 
dispersion profile of the best model compared to the observed radially-binned profile.
We followed \cite{Pryor93} to compute the observed line-of-sight velocity dispersion
profile, grouping the kinematic sample in 12 different radial bins, each containing 
250 stars, except for the last bin which has 140 stars. In the calculation of the 
observed line-of-sight velocity profile we accounted for contamination by field stars 
as in equation (\ref{for:Lv}), using the same MW Besan\c{c}on model as in Section 
\ref{subsec:dataset}.  The projected stellar number density profile is extremely well 
reproduced by our best model. A measure of the goodness of the fit to the projected surface 
density is given by the term $\ln \Ln$ of equation (\ref{for:logl}): for the best model
$\ln \Ln \simeq -30$. For comparison, for the best-fitting \cite{Sersic1968} profile of 
Fornax (\citealt{Battaglia2006}), $\ln\Ln \simeq -62.79$.  Even accounting for the different
numbers of free parameters as in equation (\ref{for:AIC}), our model gives a better 
description of the projected number density than the S\'ersic fit. This feature shows that 
our stellar DF is extremely flexible and well suited to describe the structural properties
of dSphs. Our best model has a line-of-sight velocity dispersion profile slightly increasing
with radius, which provides a good description of the observed profile. However, we recall
that in the determination of the best model we do not consider the binned line-of-sight 
velocity dispersion profile, but compare individual star data with model LOSVDs, so to 
fully exploit the available data.

Fig.~\ref{fig:best_out} plots the stellar and dark-matter density distributions, the stellar 
and dark-matter mass profiles, and the stellar anisotropy parameter profile of the best FnxCore3 model. 
The anisotropy parameter is

\begin{equation}\label{for:beta}
 \beta = 1 - \frac{\sigma_{t}^2}{2\sigma_r^2},
\end{equation}
where $\sigma_r$ and $\sigma_t$ are, respectively, the radial and tangential components of the 
velocity dispersion ($\sigma^2_t=\sigma^2_{\theta}+\sigma^2_{\phi}$, where $\sigma_{\theta}$ and 
$\sigma_{\phi}$ are angular components of the velocity dispersion in spherical coordinates; equation 
\ref{for:sigij}). Models are isotropic when $\beta = 0$, tangentially biased when $\beta < 0$ and 
radially biased when $0 < \beta \leq 1$.  The best model predicts Fornax to have slightly radially
anisotropic velocity distribution: for instance, at $r = 1\,$kpc the anisotropy parameter is
$\beta|_{1 {\rm kpc}}=0.08^{+0.14}_{-0.12}$ (see Fig.~\ref{fig:best_out}c). In our best model, the
dark matter dominates the stellar component at all radii. The dark-matter to stellar mass ratio 
is $\MratRe = 9.6^{+0.6}_{-5.7} $ within $\Reff$ and $\Mrat =  144^{+2}_{-87}$ within 3 kpc. 
The best model has a total stellar mass $\Mstot = 10^7 M_{\odot}$, which is the lower limit imposed to 
the stellar mass on the basis of observational estimates (see Section ~\ref{sec:res}). 

Fig.~\ref{fig:best_vel} compares the observed LOSVD with the LOSVD of the best model. For this 
figure the observed LOSVD was computed in the same radial bins as the line-of-sight velocity
dispersion profile of Fig.~\ref{fig:best_obs}c, while the model LOSVD is evaluated at the average 
radius of each bin: for clarity, we show only 6 of the 12 radial bins, covering the whole radial
extent of the kinematic sample. The best model has a sharply peaked LOSVD, indicative of radially
biased velocity distribution, consistent with the observed LOSVD. The contamination from MW 
field stars grows with distance from the galaxy's centre and is clearly visible in the outermost
bin. The shape of the LOSVD can be quantified by the kurtosis

\begin{equation}
 y(R) \equiv \frac{\int_{-\infty}^{+\infty} {\mathscr L}_{\star}(R, v_{||})(v_{||} - \bar{v})^4\text{d}v_{||}}
 {\left[\int_{-\infty}^{\infty} {\mathscr L}_{\star}(R, v_{||})(v_{||} - \bar{v})^2\text{d}v_{||}\right]^2},
\end{equation}
which is the fourth centred moment of the LOSVD. The bottom panel of Fig \ref{fig:best_vel} plots 
the kurtosis of the LOSVD of the best model as a function of the distance from the centre. The best 
model has a kurtosis which is constantly greater than $y=3$ (the kurtosis of a Gaussian distribution), 
which is a signature of peaked LOSVD and radial bias. 

\begin{figure}
 \includegraphics[width=1.\hsize]{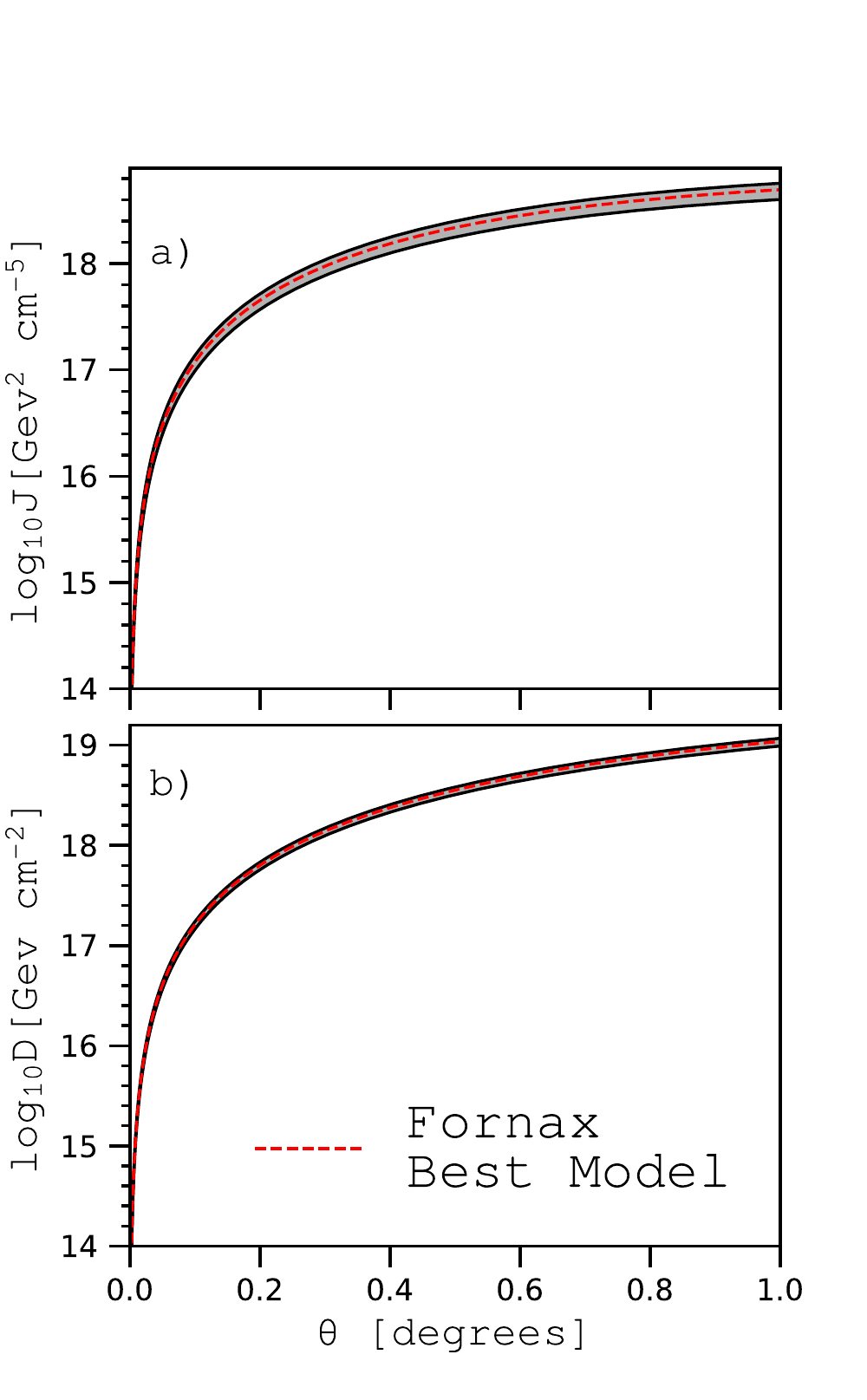} 
 \caption{Panel a: dark-matter annihilation $J$-factor (equation \ref{for:J}) of the best model of 
 Fornax (FnxCore3, dashed line) as function of the angular distance from the centre. Panel b: same
 as panel a, but for the dark-matter decay $D$-factor (equation \ref{for:D}). Bands mark the $1\sigma$ 
 uncertainty (see Section \ref{subsec:models}). }\label{fig:jdfactor}
\end{figure}

As is well known, dSphs are good candidates for indirect detection of dark-matter particles. 
The $\gamma$-ray flux due to dark-matter annihilation and decay depend on the dark-matter 
distribution through, respectively, the so called $J$ and $D$-factors. For sufficiently distant, 
spherically symmetric targets, it can be shown that the $J$-factor reduces to the integral 

\begin{equation}\label{for:J}
 J(\theta) = \frac{2\pi}{d^2}\int_{-\infty}^{+\infty}\text{d}z\int_0^{\theta d}\rhodm^2R\text{d}R,
\end{equation}
while the D-factor to

\begin{equation}\label{for:D}
 D(\theta) = \frac{2\pi}{d^2}\int_{-\infty}^{+\infty}\text{d}z\int_0^{\theta d}\rhodm R\text{d}R,
\end{equation}
where $\theta=R/d$ is the angular distance from the centre of the galaxy, $z$ is the line-of-sight and 
$d$ is the distance of the galaxy (Table \ref{tab:FnxParam}).
~Fig. \ref{fig:jdfactor} plots the $J$-factor (panel a) and $D$-factor (panel b) as functions of the 
angular distance $\theta$ computed for our best model of Fornax.
We measure at an angular distance $\theta = 0.5^{\circ}$ (corresponding approximately to the angular
resolution of the Fermi-LAT telescope in the GeV range)

\begin{equation}
 \log_{10}(J\text{ }[\text{GeV}^2\text{cm}^{-5}])=18.34^{+0.06}_{-0.09}
\end{equation}
and

\begin{equation}
\log_{10}(D\text{ }[\text{GeV cm}^{-2}])=18.55^{+0.03}_{-0.05},  
\end{equation}
consistent with \cite{Evans2016}.

\subsubsection{Performances of other families of two-component models}
\label{subsec:twocFnx}

\begin{figure*}
 \centering
 \includegraphics[width=1.\hsize]{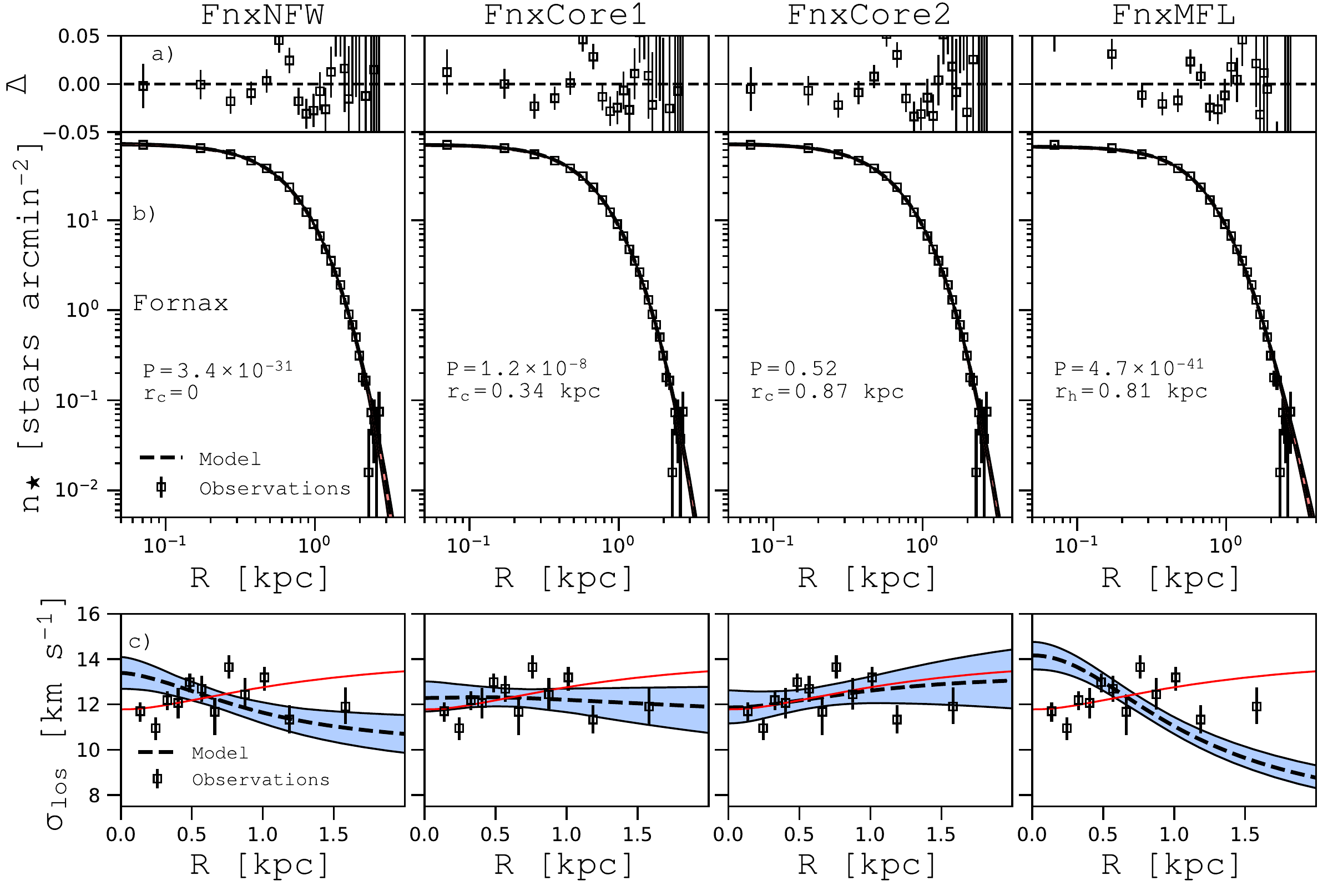}
 \caption{Columns, from left to right, refer to the best models of the FnxNFW, FnxCore1, FnxCore2 and 
 FnxMFL families, respectively. Top row of panels (a): residuals between the model and the observed 
 projected stellar density profile (points with error bars). Residuals are defined as $\Delta\equiv(\nobs-\nmod)/\nmod$.
 Middle row of panels (b): projected number density profile of the model (dashed lines), compared with
 the observed profile (points with error bars). Bottom row of panels (c): line-of-sight 
 velocity dispersion profile of the model (dashed lines), compared  with the observed
 profile (points with error bars). In panels b and c the bands indicate the 1$\sigma$ 
 uncertainties (see Section \ref{subsec:models}). In panel c the red curve shows 
 the line-of-sight velocity dispersion of the best model of any family (FnxCore3). $\rc$ 
 is the size of the core radius, $\rh$ is the stellar half-mass radius and $P$ is the
 probability of the model compared to FnxCore3 (equation \ref{for:prob}).}\label{fig:outplot2D}
\end{figure*}

\begin{figure*}
 \centering
 \includegraphics[width=1.\hsize]{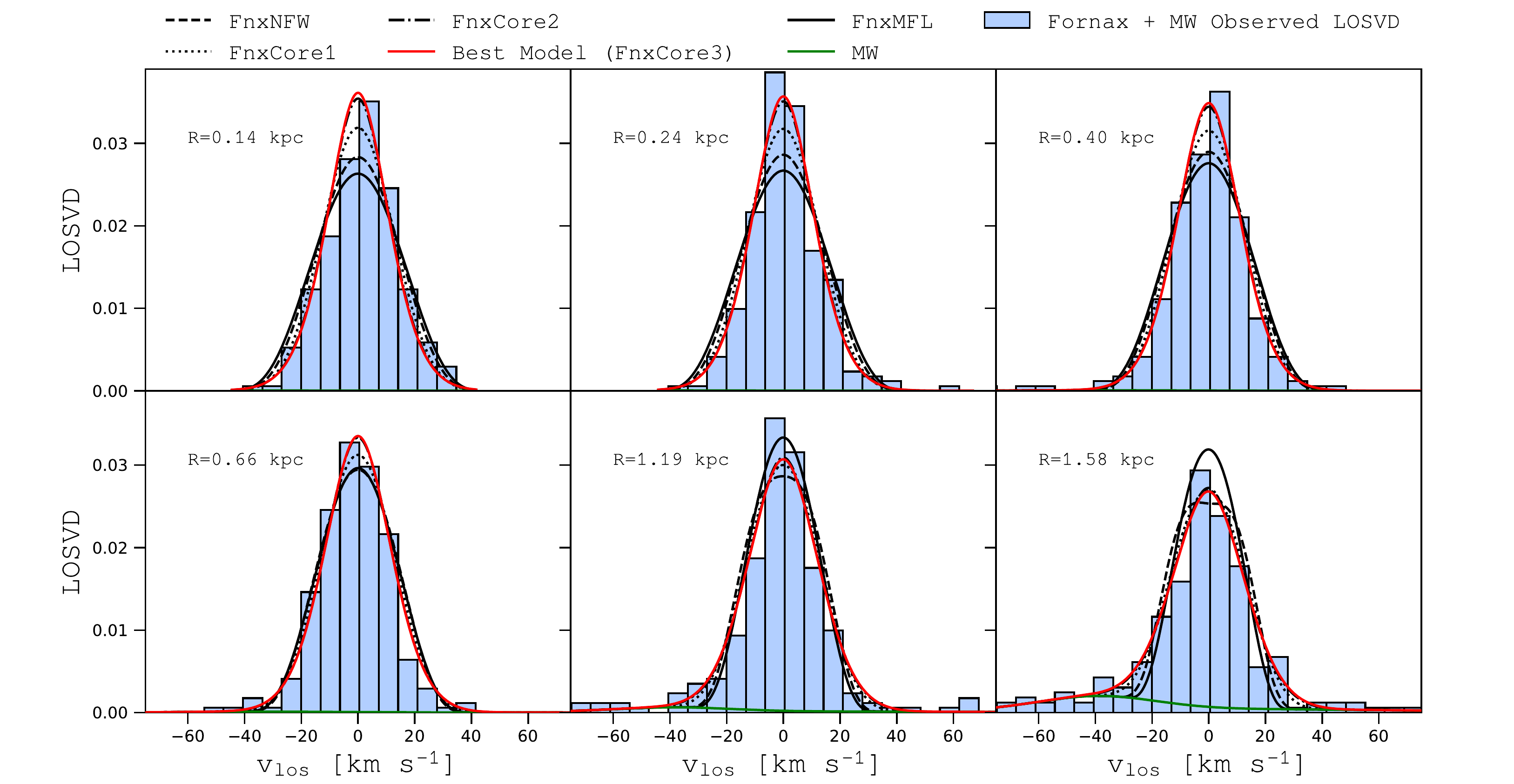}
 \caption{Comparison of the observed Fornax + MW LOSVD (histograms) and the LOSVDs of 
 the best models in the families FnxNFW, FnxCore1, FnxCore2, FnxCore3 and FnxMFL (respectively 
 dashed, dotted, dot-dashed, red solid and black solid curves). The overall best model is FnxCore3. 
 In each panel, $R$ indicates the average radius of the radial bin for which data are shown, which 
 is also the radius at which  the model LOSVD was computed. The radial bins are the same as in 
 Figs.~\ref{fig:best_vel} and \ref{fig:outplot2D}c. The green curve marks the MW's contribution.}\label{fig:velDis}
\end{figure*}

\begin{figure}
 \centering
 \includegraphics[width=1.\hsize]{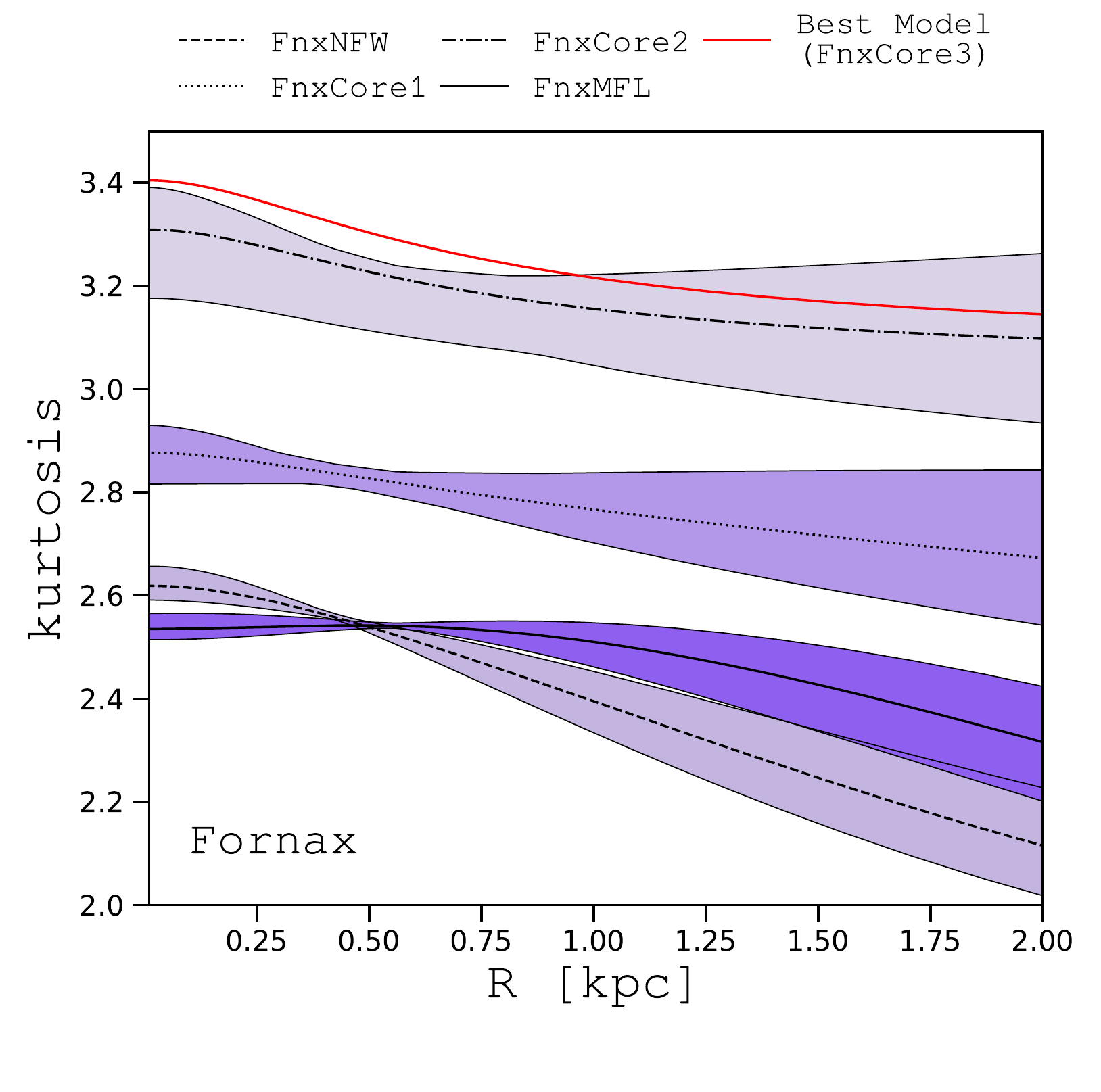}
 \caption{Kurtosis profile of the LOSVD for the best models of the families FnxNFW, FnxCore1, 
 FnxCore2, FnxMFL (dashed, dotted, dot-dashed, solid, respectively). The red curve without a 
 band shows the kurtosis profile of the best of all models (FnxCore3). The bands show the 
 1$\sigma$ uncertainties (see Section \ref{subsec:models}).}\label{fig:kurt}
\end{figure}

Here we compare the best model of Section \ref{subsec:twocb} with other families of two-component
models of Fornax. The projected number density profiles of the best models of the FnxNFW, FnxCore1, FnxCore2 
families and the observed Fornax surface density profile, and the residuals between models and 
data are plotted in Fig.s \ref{fig:outplot2D}b and \ref{fig:outplot2D}a. Fig.~\ref{fig:outplot2D}c 
shows the comparison with the line-of-sight velocity dispersion profiles. The projected number 
density profile is also well described by the other families, which have $ -40 \lesssim \ln\Ln \lesssim -25$, 
substantially better than the best-fitting S\'ersic model. Among our models, those with cored halo
reproduce well the flat behavior of the line-of-sight velocity dispersion profile, while the best 
FnxNFW predicts a slightly decreasing profile, which poorly represents the available data.  

Fig.~\ref{fig:velDis} shows the observed LOSVD compared to the model LOSVDs.
The observed LOSVD is computed in the same radial bins as in Fig.~\ref{fig:best_vel}. The
LOSVD of FnxNFW is systematically more flat-topped than that observed or the LOSVDs of
cored models, and, in the outermost bin, it has a double-peaked shape, indicative of
tangential bias. In contrast, the more extended the core of a dark-matter density
distribution, the more sharp-peaked the LOSVD is, and the more satisfying a description
of the observed LOSVD it provides (Fig.~\ref{fig:velDis}). A quantitative measure of the
shapes of a LOSVD is the kurtosis, which is plotted as a function of radius in Fig.~\ref{fig:kurt}. 
The best model of the FnxNFW family has a kurtosis which is everywhere much less than $y=3$,
while the cored families with the most extended cores have $y>3$. In other words, a 
model with NFW halo cannot reproduce at same time the flat line-of-sight velocity dispersion 
profile and the peaked LOSVD observed in Fornax.  

\begin{figure*}
 \centering
 \includegraphics[width=1.\hsize]{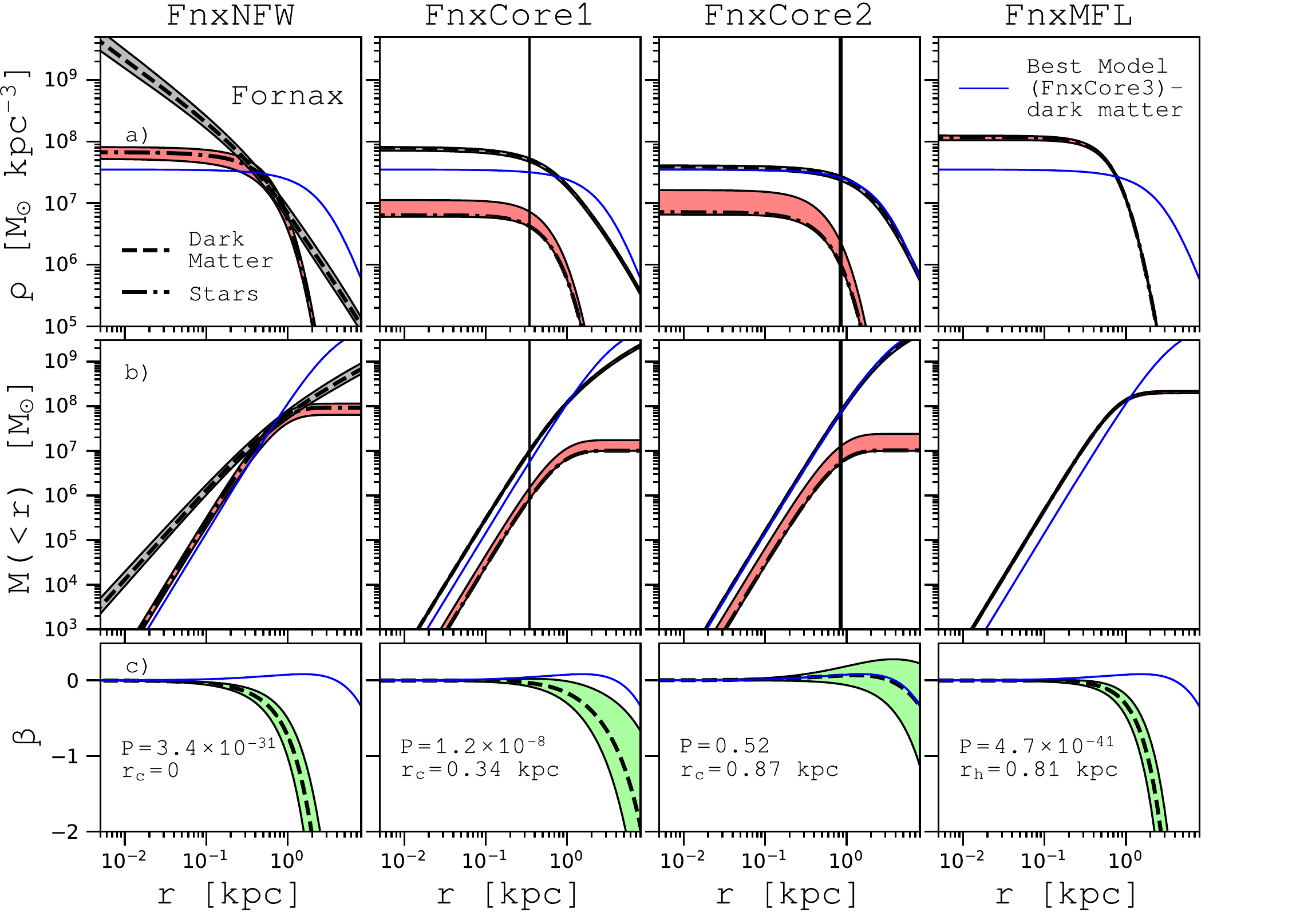}
 \caption{Columns, from left to right, refer to the best model of the FnxNFW, FnxCore1, FnxCore2
 and FnxMFL families, respectively. Top row of panels (a): stellar (dash-dotted line) and dark-matter (dashed 
 line) density profiles. Middle row of panels (b): stellar (dash-dotted line) and dark-matter (dashed line)
 mass profiles. Bottom row of panels (c): anisotropy parameter profile (dashed line). The vertical black lines
 in the four top and middle panels mark the dark halo's core radius $\rc$. In all panels the bands
 around the best fits indicate 1$\sigma$ uncertainty (see Section \ref{subsec:models}). The blue curve
 marks the dark-matter density (panels a), dark-matter mass (panels b) and stellar anisotropy profiles (panels 
 c) of the best of all models (FnxCore3). $\rh$ is is the stellar half-mass radius and $P$ is the probability
 of the model compared to the best of all models (FnxCore3).}\label{fig:outplot3D}
\end{figure*}

Figs.~\ref{fig:outplot3D}a and \ref{fig:outplot3D}b plot the stellar and dark matter
density and mass profiles, respectively.  The best models of all families with cored
halos have a total stellar mass of $10^7 M_{\odot}$, while the best NFW model has
a total stellar mass of $9.23^{+0.77}_{-2.85}\times10^7$ $M_{\odot}$. Stars never 
dominate over the dark matter in the case of cored halos, where $\MratRe = 13.4^{+0.1}_{-5.8}$ 
and $9.7^{+0.4}_{-5.7}$, respectively, for the FnxCore1 and FnxCore2 cases, whereas 
they do in the cuspy halo one, where $\MratRe = 1.12^{+0.86}_{-0.32}$. We also find a 
slight trend of the core size to be larger when the dynamical-to-stellar mass ratios are smaller.

Fig.~\ref{fig:outplot3D}c plots the profile of the stellar anisotropy parameter for the best
model in each family. It shows that the anisotropy varies with the size of the core: the more 
extended the core, the more radially biased the galaxy. Indeed, the NFW halo requires a 
highly-tangentially biased system ($\beta|_{1 {\rm kpc}}= -0.73^{+0.23}_{-0.29}$), the FnxCore1 
model requires isotropic to slightly tangential bias, while the best model, with the most extended 
core, has a preference for radial orbits (Fig.s\ \ref{fig:best_vel}, \ref{fig:velDis} and 
\ref{fig:kurt}, Table \ref{tab:bestoutp}).

By comparing the AICs (Table \ref{tab:bestoutp}), we note that, while the best model FnxCore2 is 
comparable to the best model (FnxCore3), with probability $P = 0.52$ (equation \ref{for:prob}), the 
FnxCore3 model is significantly preferable to both a model with a NFW dark halo and a model with a
small core in the dark-matter density distribution. For the FnxNFW, $\Delta$AIC$ = 140.3 $, while 
for the FnxCore1 $\Delta$AIC$ = 36.5$, values that translate in extremely small probabilities $P$
($P\simeq 3.4\times10^{-31} $ and $P\simeq10^{-8}$, respectively). We pointed out that different 
families are almost equivalent in describing the projected number density profile, so we can safely
state that most of the differences that allow us to discriminate between cored and cuspy models are 
attributable to our kinematic analysis, which minimises any loss of information (e.g. self-consistent 
LOSVD, no binning).

The best Fornax model belongs to the family with the largest core among those considered so far, 
so it is worth asking if the data allow us to put an upper limit on the dark-matter core radius $\rct$. To do that,
we run two additional experiments, considering families with core radii, respectively, 
$\rct \simeq 2.4$ ($\rc \simeq 1.94$ kpc) and $\rct \simeq 4.8$ ($\rc \simeq 3.89$ kpc). We find 
that these families have, respectively, 
$\ln \Ltot_{\rm max}=-12513.4$ and $\ln \Ltot_{\rm max} = -12514.8$, and probabilities (equation \ref{for:prob}) 
$P=0.25$ and $P=0.06$, relative to the best of all models ($\rc\simeq 1.03$ kpc). The results of these experiment suggest 
that the core of Fornax dark halo is smaller than the truncation radius ($\approx$ 2 kpc; see Section~\ref{subsec:insens}) 
of the stellar distribution.

\subsubsection{Performance of one-component models}
\label{subsec:onec}

Given that in the best two-component model (FnxCore3) the central slopes of the 
stellar and dark-matter distributions are similar (Fig.~\ref{fig:best_out}), it is 
worth exploring also a simpler one-component family of $f({\bf J})$ models. In 
particular, here we consider the case in which the only component has the DF given
by equation (\ref{for:df1}). This family of models can be interpreted as describing 
a system without dark matter, but also as mass-follows-light (MFL) models, in which dark
matter and stars have the same distribution. We will refer to this family of models as
FnxMFL. Since in this case $\Mfdm=0$, this family has four free parameters ($\alpha$, 
$\eta$, $\Jfst$, $\Mfst$; equation \ref{for:df1}). In Tab. \ref{tab:bestParam} we report
the parameters corresponding to the best FnxMFL model. The right column of Fig.~\ref{fig:outplot2D} 
plots the projected number density profile and the line-of-sight velocity profile 
of the best FnxMFL model. The projected number density profile is well reproduced 
also by the MFL models, for which $\ln \Ln \simeq -40$, still much better than
a S\'ersic fit, while the line-of-sight velocity dispersion profile is clearly 
far from giving a good description of the observed profile. Fornax MFL
models are rejected with high significance: we find $\Delta$AIC$ = 185.74$, the 
largest $\Delta$AIC among our models, consequently, with a probability $P\simeq10^{-41}$.
 
In Fig.~\ref{fig:velDis} the LOSVD of the FnxMFL model is compared with the LOSVD of the 
two-component models. MFL models tend to underestimate the observed LOSVD in the innermost regions
(top three panels) and to overestimate it in the outermost regions (bottom three panels).

The rightmost column of Fig.~\ref{fig:outplot3D} plots in panels a, b and c, respectively, the 
density, mass and anisotropy parameter profile predicted by the best FnxMFL model, which has total
mass $\Mstot = 2.06^{+0.13}_{-0.12}\times10^8$ $M_{\odot}$. Under the assumption 
that the dark halo follows the density distribution of the stellar component, this value is an
indication of the dynamical (stellar plus dark-matter) mass. The FnxMFL model is tangentially
anisotropic with $\beta|_{1 {\rm kpc}} = -0.32^{+0.13}_{-0.16}$. The main parameters of this model 
are summarised in Tables \ref{tab:bestParam} and \ref{tab:bestoutp}.

\subsubsection{Insensitivity to the halo scale radius}
\label{subsec:insens}

All the two-component models considered above have the scale radius of the dark halo fixed to 
$\rst = 4$. In this Section we relax this assumption and let $\rst$ vary. Of course 
we are interested only in exploring cosmologically motivated values of $\rst$, which can be 
evaluated as follows. According to current estimates of the low-mass end of the stellar-to-halo mass relation 
(\citealt{Read2017}), galaxies with stellar mass $\Mstot = 10^7 - 10^8$ $M_{\odot}$ (such
as Fornax) have virial mass $4.5 \times 10^9 \lesssim \Mvir \lesssim 3 \times 10^{10}$ $M_{\odot}$
and virial radius\footnote{The dark halos of satellite galaxies
such as Fornax are expected to be tidally truncated 
at radii much smaller than $\rvir$. In this context the value of $\rvir$ expected in the absence of truncation is 
used only as a reference to estimate $\rs$.} $35 \lesssim \rvir/$kpc $\lesssim 61$. According to the halo mass-concentration relation 
(\citealt{Cuartas2010}), in the present-day Universe halos in this mass range have $14\lesssim \rvir/\rs \lesssim 16$,
so $2\lesssim\rs/$kpc$\lesssim5$, or $2.5\lesssim \rst \lesssim6.2$, for $\rh \simeq 0.81$ kpc, which 
is the stellar half-mass radius of Fornax. 

Even the lower bound of this cosmologically motivated interval of values of the scale radius ($\rs \simeq 2$
kpc) is beyond the truncation of the stellar component of Fornax (97\% of the stellar mass is 
contained within 2 kpc; see Fig.s \ref{fig:best_out}b and \ref{fig:outplot3D}b), so we do expect 
our results to be insensitive to the exact value of $\rs$ within the above
range. However, given the very poor performance of the NFW models in reproducing the observed 
kinematics of Fornax (Section \ref{subsec:twocFnx}), we explored also a more general family of NFW 
models, named FnxNFW-rs, in which $\rst$ is a free parameter, in the range $2.5 - 6.2$.
As predicted, these models turned out to be poorly sensitive to $\rs$, with a slight preference for higher values. 
The best model of this new NFW family has $\rst = 6.04^{+0.16}_{-3.52}$, so all 
the explored values of $\rst$ are within $1\sigma$. This model has $\ln \Ltot =-12581.14$ and AIC$ = 25174.28$ 
(see Table \ref{tab:fornaxrs}), which, compared to the best model (FnxCore3), gives $\Delta$AIC$ 
\sim 140.26$, approximately the same $\Delta$AIC as the best model of the family FnxNFW (Section \ref{subsec:twocFnx}).
We conclude that the results obtained fixing $\rst$ are robust against 
uncertainties on this parameter.

\begin{table*}
\begin{center}
\renewcommand{\arraystretch}{1.8}
\begin{tabular}{p{1.7cm}cccccc}
\hline\hline
 Family		& $\alpha$	& $\eta$	 & $\Mftil$	&	$\Jftil$ 	&	$\Jfst$/[km s$^{-1}$ kpc]	&	$\Mfst$/[$M_{\odot}$]  \\
 \hline\hline
 FnxNFW-rs	& $1.36^{+0.03}_{-0.04}$ & $0.37^{+0.03}_{-0.02}$ & $439^{+87}_{-182}$ & $100.95^{+7.10}_{-11.25}$ & $4.75^{+0.46}_{-0.30}$ & $1.30^{+0.66}_{-0.17}\times10^7$ \\ 
 \hline\hline
		& $\rst$	&	$\Mdmt$ 	 & $\ln \Ltot_{\rm max}$ & ${\rm AIC}$	& $\Delta$AIC &	$P$ \\
 \hline\hline
 FnxNFW-rs	& $6.04^{+0.16}_{-3.52}$ & $110.9^{+19.5}_{-34.9}$  &-12581.14081 & 25174.28 & 140.25 & $3.5\times10^{-31}$ \\ 
 \hline\hline
\end{tabular}
\caption{Parameters of the best Fornax model of the FnxNFW-rs family with free scale radius (Section \ref{subsec:insens}).
$\alpha$ and $\eta$: parameters of the stellar DF (\ref{for:df1}). $\Mftil \equiv \Mfdm/\Mfst$. $\Jftil\equiv\Jfdm/\Jfst$. 
$\Jfst$ and $\Mfst$: respectively, action and mass scales (equation \ref{for:df1}). $\Mdmt \equiv \Mdmt/\Mstot$. 
As in the family FnxNFW (Table \ref{tab:bestParam}) $\Jctil\equiv\Jcdm/\Jfdm=0$ and $\Jttil\equiv\Jtdm/\Jfdm=6$.
$\Mfdm$, $\Jfdm$, $\Jcdm$ and $\Jtdm$ are the parameters of the dark-matter DF (equations \ref{for:dmdf}-\ref{for:expcoffDF}). 
$\rst\equiv\rs/\rh$. $\Mdmt\equiv \Mdmtot/\Mstot$. $\rs$ and $\rh$ are, respectively 
the halo scale radius and the half-mass radius of the stellar component; $\Mdmtot$ and $\Mstot$ are, 
respectively, the total dark-matter and stellar masses (equations \ref{for:totmassDM} and \ref{for:totmass}). 
$\ln \Ltot_{\rm max}$: log-likelihood (equation \ref{for:logl}). ${\rm AIC}$: value of 
the Akaike Information Criterion (equation \ref{for:AIC}). $\Delta$AIC: difference between the AIC of the 
FnxNFW-rs and the best of all models (FnxCore3, see Table \ref{tab:bestoutp}). $P$: probability 
that the FnxNFW-rs best model represents the data as well as the best of all models (FnxCore3).}\label{tab:fornaxrs}
\end{center}
\end{table*}


\subsection{Comparison with previous work}
\label{subsec:comp}

\begin{figure}
 \centering
 \includegraphics[width=1.\hsize]{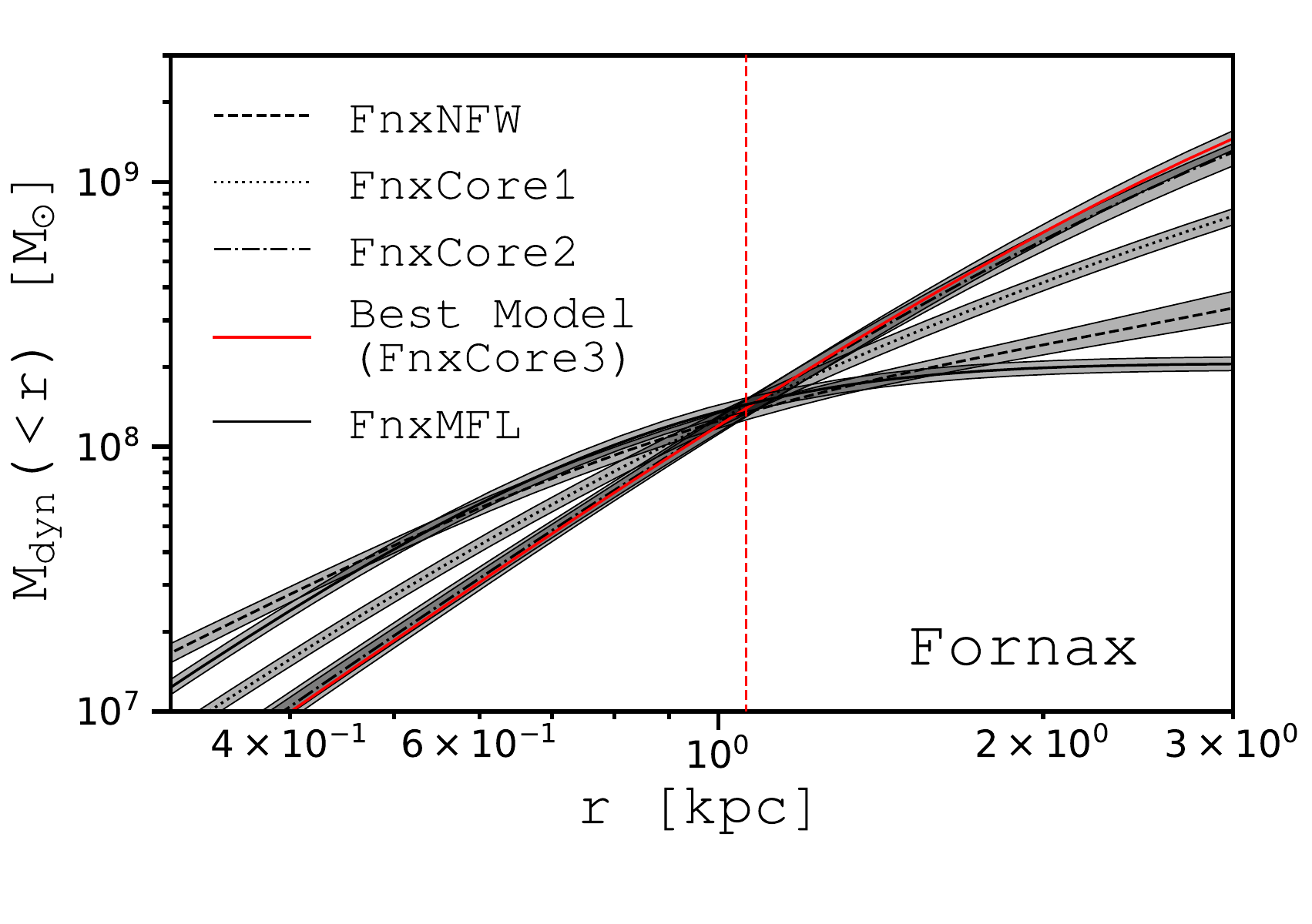}
 \caption{Total mass profiles (stars and dark matter) of the FnxNFW, FnxCore1, FnxCore2, FnxCore3
 and FnxMFL families. The bands mark the 1$\sigma$ uncertainty (see Section
 \ref{subsec:models}). The vertical red-dashed line indicates $r_{\rm m}$, radius where the total
 mass is model-independent.}\label{fig:masstot}
\end{figure}

Here we compare the results of our dynamical modelling of Fornax with previous works. 
Fig.~\ref{fig:masstot} plots the dynamical (stars plus dark matter) mass profile of the best
of our models (FnxCore3) compared to those of the best models of other families. Within the
radius $r_{\rm m} \simeq 1.7$ $\Reff$ $\simeq 1.05$ kpc, the dynamical mass is robustly
constrained against changes in the specific shape of the dark halo and the anisotropy.
In our best model, the total mass enclosed within $r_{\rm m}$ is $\Mdn(r_{\rm m}) =
1.38^{+0.10}_{-0.10} \times 10^8 M_{\odot}$, consistent with the mass estimate of
\cite{AmoriscoEvans2011} of $\Mdn(1.7$ $\Reff) \simeq 1.3\times10^8$ $M_{\odot}$.
\cite{AmoriscoEvans2011} performed a dynamical study of 28 dSphs, using different
halos and modelling the stellar component with an ergodic King DF (\citealt{Michie1963}, 
\citealt{King1966}). Remarkably, they find that, for all the dSphs in their sample, 
the best mass constraint is given at $r_{\rm m} \simeq 1.7$ $\Reff$.

\cite{Strigari2008} performed a Jeans analysis on a sample of 18 dSphs. They used analytical 
density distributions for the dark matter in order to describe both cuspy and cored models,
and studied the cases of anisotropic stellar velocity distributions, with radially varying 
anisotropy. They use a maximum likelihood criterion based on individual star velocities, 
assuming Gaussian LOSVDs. For all the dSphs, the authors find that $\Mdn(300$ pc$)$, the 
total mass within 300 pc, is well constrained, and they estimate for Fornax $\Mdn(300$ pc$)
= 1.14^{+0.09}_{-0.12}\times10^7$ $M_{\odot}$, For our best model we find a smaller value, 
$\Mdn(300$ pc$)=0.44^{+0.07}_{-0.03}\times 10^7$ $M_{\odot}$.

\cite{Walker2009} performed a Jeans analysis on a wide sample of dSphs finding that a robust
mass constraint is given at $\Reff$, where, for the Fornax dSph, they measure $\Mdn(\Reff) =
4.3^{+0.6}_{-0.7} \times 10^7$ $M_{\odot}$, marginally consistent with $\Mdn(\Reff) =
3.37^{+0.33}_{-0.22}\times10^7$ $M_{\odot}$, that we get for our best model.


The existence of a particular radius where the total mass is tightly constrained over a wide 
range of dark halo density profiles and anisotropy has been noted by many authors 
(\citealt{Penarrubia2008}, \citealt{Strigari2008}, \citealt{Walker2009}, \citealt{Wolf2010}).
However, there is not always agreement on the value of this particular radius, so it is
worth asking why these differences arise. Dynamical modelling faces the problem that since
one has to deal with only a 3D projection of the six-dimensional phase space (two coordinates 
in the plane of the sky and the line-of-sight velocities), the DF is not fully constrained
by observations. Jeans analysis provides a work-around: the Jeans equations predict relations
between some observables without delivering the DF and they do not require significant 
computational effort.  However, Jeans analysis is not conclusive, because it is not
guaranteed that the resulting model is physical in the sense that it has an everywhere 
non-negative DF (e.g. \citealt{CiottiMorganti2010}, \citealt{AmoriscoEvans2011}). Moreover, 
it involves differentiation of the data and does not deliver the LOSVD but only its first two 
moments. By contrast, the non-negativity of all our DFs is guaranteed, our procedure does
not entail differentiation of the data, and we can exploit all the information that is 
contained in the LOSVD.  It is reassuring that our estimate of $\Mdn(1.7 \Reff)$ is consistent
with \cite{AmoriscoEvans2011}, which is, to our knowledge, the only other work in which Fornax 
is modeled starting from DFs.

\begin{figure*}
 \centering
 \includegraphics[width=1.\hsize]{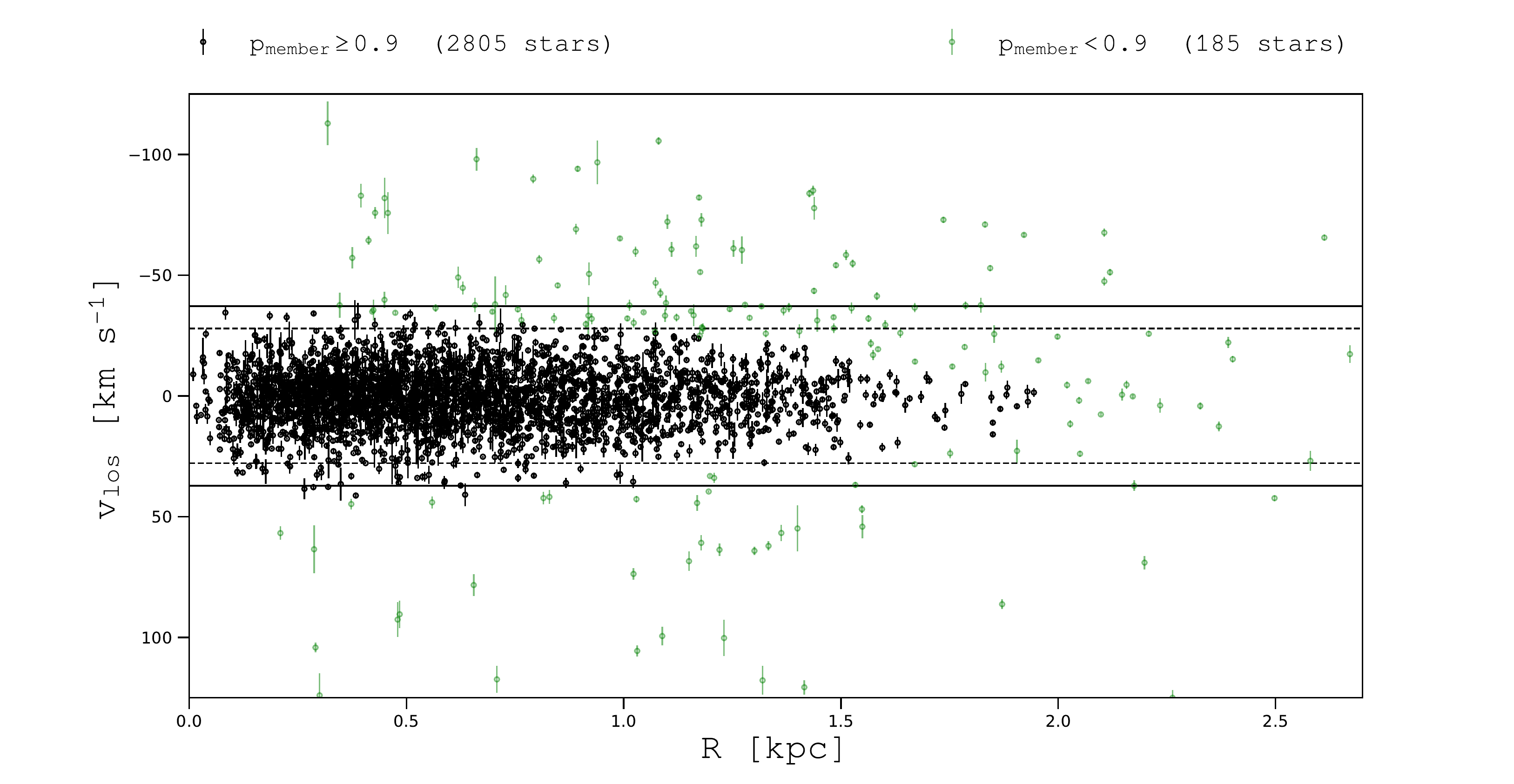}
 \caption{Position-velocity diagram of the Fornax kinematic sample. Different colours mark 
 different membership  probabilities $p_{\rm member}$; the horizontal dashed and solid lines 
 mark the regions obtained by using an iterative $2.5\sigma$-clip and $3\sigma$-clip, 
 respectively.}\label{fig:veldismemb}
\end{figure*}

Recently, \cite{Diakogiannis2017} presented a new, spherical, non-parametric Jeans 
mass modelling method, based on the approximation of the radial and tangential components 
of the velocity dispersion tensor via $B$-splines and applied it to the Fornax dSph.
Even considering different cases of dark-matter density distributions, they find 
that the best Fornax model is a simple MFL model. In our case, the MFL scenario is
rejected with high significance (see Table \ref{tab:bestoutp}). The authors measure a
total mass of $\Mdn = 1.613^{+0.050}_{-0.075}\times 10^8$ $M_{\odot}$, which is slightly smaller 
than the total mass of our MFL models, $2.06^{+0.13}_{-0.12}\times10^8$
(see Section \ref{subsec:onec}). The best model of \cite{Diakogiannis2017} 
is characterised by tangential anisotropy, with mean anisotropy $\langle \beta \rangle = 
-0.95^{+0.78}_{-0.72}$, in agreement with the values we obtain from  our FnxMFL models,
which predict Fornax to be tangentially biased, with a reference anisotropy $\beta|_{1 {\rm kpc}}
=-0.32^{+0.13}_{-0.16}$. There are several differences between our analysis and that 
of \cite{Diakogiannis2017} that together explain the different conclusions about MFL
models of Fornax. We believe that our model-data comparison is more accurate in some
respects, which makes our conclusions more robust. For instance, we use a more extended
observed stellar surface density profile and we account self-consistently for the MW 
contamination.

\cite{BreddelsHelmi2013} applied spherical \cite{Schwarzschild1979} modelling to four of
the classical dSphs, including Fornax, assuming NFW, cored and \cite{Einasto1965} dark-matter
density profiles. They use both the second and the fourth moment of the LOSVD in comparisons
with data. They conclude that models with cored and cuspy halo yield comparable fits to the 
data, and they find that models conspire to constrain the total mass within $1\,$kpc to a 
value $\Mdn(1$ kpc$) \simeq 10^8$ $M_{\odot}$ that is in good agreement with our value, 
$\Mdn(1$ kpc$)=1.20^{+0.09}_{-0.08} \times10^8$ (Fig.~\ref{fig:masstot}). \cite{BreddelsHelmi2013}
find that the data for Fornax are consistent with an almost constant, isotropic or slightly 
tangential-biased anisotropy parameter profile $\beta = -0.2\pm0.2$, marginally consistent
with our almost isotropic values.

As far as the central dark-matter distribution is concerned, our results confirm and strengthen
previous indications that Fornax has a cored dark halo. For instance, \cite{Goerdt2006} argue 
that the existence of five globular clusters in Fornax is inconsistent with the hypothesis of a
cuspy halo since, due to dynamical friction, the globular clusters would have sunk into the 
centre of Fornax in a relatively short time (see also \citealt{Sanchez2006}, \citealt{ArcaSedda2016}).
\cite{Amorisco2013}, exploiting the information
on the spatial and velocity distributions of Fornax subpopulations of stars, showed that a
cored dark halo represents the data better and were able to constrain the size of the core, 
finding $\rc = 1^{+0.8}_{-0.4}\,$kpc, which agrees with the size of the core of our best model.
\cite{Jardel2012} applied Schwarzschild axisymmetric mass models to Fornax, testing NFW and
cored models with and without a central black hole. They used the LOSVD computed in radial bins
to constrain the models, finding that the best model has a cored dark halo. They also computed 
the anisotropy profile according to their best model selection and argue that Fornax has a 
slightly radially biased orbit distribution, in agreement with our estimate.
\cite{Walker2011}, considering two different stellar subpopulations
of Fornax, provided anisotropy-independent estimates of the enclosed mass within 560 pc and 900
pc, $M(560$ pc$) = 3.2\times10^7 M_{\odot}$ and $M(900$ pc$) = 11.1\times10^7 M_{\odot}$, which
are in perfect agreement with our results (Fig.~\ref{fig:masstot}).

\subsection{Membership}
\label{subsec:memb}

As a further application of our DF-based method, we computed the probability that each star 
of the kinematic sample of Fornax is a member of the dSph. Contaminants are objects that, due
to projection effects, seem to belong to an astrophysical target, but that are intrinsically
located in foreground or background.  Separating member stars from foreground contaminants 
is not an easy task, especially when they have similar magnitudes, colours, metalliticies,
or when foreground stars move at similar velocities with respect to the target's systemic 
velocity: this is, in particular, the case for Fornax. This makes usual approaches, such as 
the $n\sigma$-clip of the line-of-sight velocity of stars, ineffective. The $n\sigma$-clip 
strongly depends on the choice of the threshold $n$ and, in cases such as that of Fornax,
it does not ensure the reliable exclusion of contaminants.  Thus, we use an alternative approach
to define a posteriori membership probabilities that relies on the LOSVD of our best model and
of the Besan\c{c}on model of the foreground. 

We define $p_{\rm member}$ the probability that a star belongs to a certain target (in our
case Fornax) and $p_{\rm cont} \equiv 1 - p_{\rm member}$ the probability that the stars belongs
to the contaminants population. In general

\begin{equation}
 p_{\rm member} \equiv p_{\rm member}({\boldsymbol \theta}),
\end{equation}
where ${\boldsymbol \theta}$ describes some measured properties of the stars. Let us focus on the
simple case in which ${\boldsymbol \theta} = (R, \vlos)$ and define the membership probability
of the $k$-th star as

\begin{equation}\label{for:member}
 p_{{\rm member}, k} = \frac{(1-\omega_k)\int_{-\infty}^{+\infty}\mathcal{L}_{\star}(R_k,v_{||})G_k(v_{||}-\vlosk)\text{d}v_{||}}
 {\int_{-\infty}^{+\infty}\mathcal{L}_{\rm tot}(R_k,v_{||})G_k(v_{||}-\vlosk)\text{d}v_{||}},
\end{equation}
where $\mathcal{L}_{\star}, \mathcal{L}_{\rm tot}, G_k$ are as in Section \ref{sec:Analysis2fj} and
are functions of $\theta$. Here $\mathcal{L}_{\star}$ is the LOSVD of the best model, while the term
$\omega_k$ is a function of $R$, controlling the relative contribution between contaminants and Fornax 
(equation \ref{for:omegak}). We account for the errors on single velocities through the convolution
with $G_k$, a Gaussian function with mean equal to the $k$-th velocity and standard deviation equal 
to $\delta\vlosk$.  Fig.~\ref{fig:veldismemb} shows the position-velocity diagram of the Fornax 
kinematic sample, where different colours mark stars with different probability of membership.  
We identify $2805$ stars with $p_{\rm member} \geq 0.9$, that can be safely interpreted as Fornax 
members, while $94$ stars have probability $p_{\rm member} < 0.1$, corresponding mostly to high-velocity 
and/or distant stars. Fig.~\ref{fig:veldismemb} shows the region delimited by selecting stars using
an iterative $n\sigma$-clip, with $n=2.5, 3$.  In the case of Fornax, a $n\sigma$-clip leads inevitably 
to the MW's contribution being underestimated, especially in the outermost regions, which are likely 
to be dominated by foreground stars, and to classification as contaminants of stars of that lie in the
innermost regions but belong to the high-velocity tail of the LOSVD. Any attempt to alleviate this 
problem by increasing the threshold $n$, would have the effect of amplifying the underestimate of
the contaminants at larger distances.

Our approach does not guarantee a perfect distinction between members and contaminants, especially 
close to $\vlos \simeq 0$, but by using a self-consistent model for the target LOSVD we maximise our
chances of selecting likely members.

\section{Conclusions}
\label{sec:conc}

We have presented new dynamical models of a dSph based on DFs depending on the action integrals. 
In particular, we combined literature DFs (\citealt{Posti2015}; \citealt{ColeBinney2017}) with a
new analytical DF to describe the stellar distribution of a dSph in both its structural and
kinematic properties. In their most general form our models make it possible to represent 
axisymmetric and possibly rotating multi-component galaxies, including the dark halo and different
stellar populations, each of which is described by a DF. The adiabatic invariance of the actions
allows us to distinguish between adiabatic contraction of the dark halo during baryon accretion 
and evolution of the dark halo arising from upscattering of dark-matter particles, whether by a 
bar, sudden ejection of mass by supernovae, or infalling satellites and gas clouds. The use
of the DFs allows us to compute the stellar LOSVD of the models, which is a key instrument 
in the application to observed dSphs. In the model-data comparison we use the velocities
of individual stars and we account for contamination by field stars.

We applied our technique to the Fornax dSph, limiting ourselves for simplicity to spherically 
symmetric models. We explored both two-component models (with both cuspy and cored
dark halos) and simpler one-component MFL models. The model that best reproduces 
Fornax observables is a model with a dark halo that has quite a large  core:
$\rc\simeq1.05\,{\rm kpc} \simeq 1.7\Reff$. We find that Fornax is everywhere dark-matter
dominated, with dark-to-luminous-mass $(\Mdm/M_{\star})|_{\Reff}= 9.6^{+0.1}_{-5.7}$ 
within the effective radius and $(\Mdm/M_{\star})|_{3\,{\rm kpc}} = 144^{+2}_{-88}$ 
within $3\,$kpc. The self-consistent stellar velocity distribution of the best model 
is slightly radially biased: the anisotropy profile is relatively flat, with $\beta=0$ 
in the centre and $\beta= 0.08^{+0.13}_{-0.12}$ at 1 kpc. Our best model is preferred 
with high statistical significance to models with a NFW halo and to MFL models; the
latter are several orders of magnitude less likely than our best model. The strength 
of this conclusion derives not only from the fact that, starting from the DFs, we 
implicitly exclude unphysical models, but also because by performing a star-by-star   
comparison with the self-consistent LOSVDs of the models, we fully exploit the available
kinematic data. For instance, our analysis demonstrates that models with cuspy NFW halos
cannot reproduce at the same time the flat line-of-sight velocity dispersion profile and the
peaked LOSVDs of Fornax. Our results confirm and strengthen previous indications that 
Fornax is embedded in a dominant cored dark halo.

A knowledge of the present-day dark-matter distributions of dSphs is important because it
has implications for both models of galaxy formation and the nature of dark matter.  
In the context of the standard $\Lambda$CDM cosmological model, the fact that Fornax today 
has a cored dark halo can be interpreted as a signature of the gravitational interaction
of gas and dark matter during galaxy formation, which modified an originally cusped halo.
In alternative dark-matter theories (e.g. the so-called fuzzy dark matter model; \citealt{Hui2017}), 
the core is an original feature of the cosmological dark halo, independent of the interaction
with baryons. Experiments trying to detect dark matter indirectly via annihilation or decay
in dSphs rely on the knowledge on the $J$-factor and the $D$-factor of these systems, which 
require accurate measures of the dark-matter distribution in the central regions of these 
galaxies. For our best model of Fornax we find $\log_{10}(J/[$GeV$^2$ cm$^{-5}])=18.34^{+0.06}_{-0.09}$ 
and $\log_{10}(D/[$GeV cm$^{-2}])=18.55^{+0.03}_{-0.05}$, for aperture radius
$\theta=0.5^{\circ}$. 

In this paper we have shown that $f({\bf J})$ DFs are powerful tools for the dynamical
modelling of dSphs. As a first application we have modeled the Fornax dSph as a two-component
(star and dark matter) spherically symmetric system. In the near future we plan to perform
similar analyses on other dSphs and to fully exploit the power of the presented method by
exploring axisymmetric models either with multiple stellar populations or
using an extended stellar DF, depending on metallicity as well as on the action integrals 
(\citealt{Das2016}).

\section*{Acknowledgments}

We thank G. Battaglia, M. Breddels and M. Walker for useful discussions and for sharing 
their data, and B. Nipoti and E. Vasiliev for helpful suggestions. JJB and AGAMA are funded by
the European Research Council under the European Union's Seventh Framework Programme 
(FP7/2007-2013)/ERC grant agreement no.~321067.

\appendix
\section{f({\bf J}) Total Mass}
\label{app:A}

Here we derive an expression of the total mass of a system described by an action-based 
DF $f({\bf J})$. Given an $f({\bf J})$ distribution function which depends on the action 
integrals through a homogeneous function $h({\bf J}) = J_r + \omega(|J_{\phi}| + J_z)$, the 
total mass $M$ of the system is given by
\begin{equation}\label{for:fjmass1}
 \frac{M}{(2\pi)^3} = \int \text{d}^3{\bf J}f({\bf J}) = \int_{-\infty}^{\infty}
 \text{d}J_{\phi}\int_{0}^{\infty}\text{d}J_z\int_{0}^{\infty}f({\bf J})\text{d}J_r.
\end{equation}
When $h({\bf J})$ is even in $J_{\phi}$ we can write equation (\ref{for:fjmass1}) as
\begin{equation}\label{for:fjmass2}
 \frac{M}{(2\pi)^3} = 2 \int_{0}^{\infty}\text{d}J_{\phi}\int_{0}^{\infty}\text{d}J_z
 \int_{0}^{\infty} f({\bf J})\text{d}J_r. 
\end{equation}
Changing coordinates from ($J_r, J_{\phi}, J_z$) to ($J_r, L, J_z$), where $L$ is the 
total angular momentum modulus, and integrating out $J_z$ ($0 < J_z < L$),
equation (\ref{for:fjmass2}) becomes
\begin{equation}\label{for:fjmass3}
\frac{M}{(2\pi)^3} = 2 \int_{0}^{\infty}\text{d}J_r\int_{0}^{\infty}Lf(J_r, L)\text{d}L.
\end{equation}
Finally, changing coordinates from ($J_r,L$) to ($L, h$) and integrating out L 
($0 < L < h/\omega$), equation (\ref{for:fjmass3}) becomes\footnote{ 
This equation was derived in \cite{Posti2015}. Note, however, that there is a typo in equation 
36 of \cite{Posti2015}.}
\begin{equation}
 \frac{M}{(2\pi)^3} = \frac{1}{\omega^2}\int_0^{\infty}h^2f(h)\text{d}h.
\end{equation}

\section{APPLICATION TO MOCK DATA}
\label{app:mock}

\begin{table*}
\begin{center}
\begin{tabular}{lccccccccccc}
 \hline\hline
 Parameter	&$m$&	$R_S$ [kpc]& $\rho_0$ [$M_{\odot}$/kpc$^{-2}]$  & $r_{\rm s, dm}$/[kpc] & $r_{\rm t, dm}$/[kpc] &$N_{\rm tot, \star}$ &$R_{\rm e}$ [kpc]& 	$\Mstot$ $[M_{\odot}]$&$N_n$&$N_v$& $n_{\rm f}$ [stars kpc$^{-2}$] \\
 \hline
 value		&0.71&	0.58	   & $4.539\times10^6$ &  4 & 20  &  51200         &0.62	     	    &		$5\times10^7$ &27   &3000 & 66.8 \\	
 \hline\hline
\end{tabular}
\caption{Main parameters adopted to generate the mock. $m$ and $R_S$: index and scale radius of the deprojected
S\'ersic profile (equation \ref{for:sersic_mock}); $\rho_0$, $r_{\rm s, dm}$ and $r_{\rm t, dm}$: respectively, 
the reference density, scale radius and truncation radius of the mock dark-matter component (equation \ref{for:dmcomp});
$N_{\rm tot}$: total number of stars; $R_{\rm e}$: effective
radius; $\Mstot$: total mass; $N_n$: number of bins of the projected stellar number density profile; $N_v$: number 
of stars of the kinematic sample; $n_{\rm f}$: mean projected density of mock field stars.}\label{tab:mockParam}
\end{center}
\end{table*}

\begin{table*}
\begin{center}
\renewcommand{\arraystretch}{1.8}
\begin{tabular}{p{1.7cm}ccccccccccc}
\hline\hline
Family		& $\alpha$	& $\eta$	&   $\Mftil$ 	&	$\Jftil$ 	&	$\Jfst$/[km s$^{-1}$ kpc]	&	$\Mfst$/[$M_{\odot}$]  &	$\Jctil$ \\
 \hline\hline
 mockNFW	& $1.48^{+0.06}_{-0.05}$ & $0.54^{+0.02}_{-0.03}$ &   $2.02^{+0.82}_{-0.87}\times10^3 $&	$84.20^{+9.06}_{-30.30}$ & $7.15^{+0.53}_{-0.55}$ & $5.38^{+1.58}_{-1.72}\times10^6$ &	0\\ 
 
 mockCore1	& $1.17^{+0.08}_{-0.04}$ & $0.72^{+0.05}_{-0.04}$ & $1.45^{+0.91}_{-0.94}\times10^3$	&   $70.84^{+7.80}_{-9.26}$	& $6.13^{+0.79}_{-0.50}$	&  $6.10^{+9.07}_{-2.17} \times10^6$& 0.02	\\
 
 mockCore2	& $1.11^{+0.06}_{-0.05}$ & $0.76^{+0.05}_{-0.03}$ & $227^{+725}_{-106}$	&  $42.70^{+17.90}_{-8.78}$	&	 $5.76^{+0.57}_{-0.50}$	&  $2.83^{+0.87}_{-0.78}\times10^7$	&	0.2\\
 \hline\hline
		&	$\rct$	&		$\Mdmt$	&$\ln \Ltot_{\rm max}$ & ${\rm AIC}$  & $\Delta$AIC & 	$P$ \\
 \hline\hline
 mockNFW	&	0	& $1300^{+391}_{-558}$ & -12249.44 & 24510.88	&	0	&	1 \\
 
 mockCore1	&	$0.414^{+0.013}_{-0.026}$	& $964^{+459}_{-613}$	&  -12271.70	&   24553.4 &	42.53 &	 $5.85\times10^{-10}$\\
 
 mockCore2	&	$0.893^{+0.144}_{-0.086}$	& $146^{+58}_{-31}$	& -12268.56	&  24547.12	&	36.24&	$1.35\times10^{-8}$ \\ 
 \hline\hline
 \end{tabular}
\caption{Parameters of the best mock models of each family. $\alpha$ and $\eta$: parameters
of the stellar DF (\ref{for:df1}). $\Mftil \equiv \Mfdm/\Mfst$. $\Jftil\equiv\Jfdm/\Jfst$. $\Jfst$
and $\Mfst$: respectively, action and mass scales (equation \ref{for:df1}). $\Jctil\equiv\Jcdm/\Jfdm$. 
All models have $\Jttil\equiv\Jtdm/\Jfdm=6$. $\Mfdm$, $\Jfdm$, $\Jcdm$ and $\Jtdm$ are the parameters
of the dark-matter DF (equations \ref{for:dmdf}-\ref{for:expcoffDF}). $\rct\equiv\rs/\rh$. $\Mdmt\equiv
\Mdmtot/\Mstot$. $\rs$ and $\rh$ are, respectively the halo scale radius and the half-mass radius of 
the stellar component; $\Mdmtot$ and $\Mstot$ are, respectively, the total dark-matter and stellar masses
(equations \ref{for:totmassDM} and \ref{for:totmass}). $\ln \Ltot_{\rm max}$:
log-likelihood (equation \ref{for:logl}). ${\rm AIC}$: value of the Akaike Information Criterion
(equation \ref{for:AIC}). $\Delta$AIC: difference between the AIC of the best model of a family and 
the AIC of the best of all models (mockNFW). $P$: probability that a model represents the data as 
well as the best of all models (mockNFW).}\label{tab:mocktab}
\end{center}
\end{table*}

\begin{figure}
 \centering
 \includegraphics[width=1.\hsize]{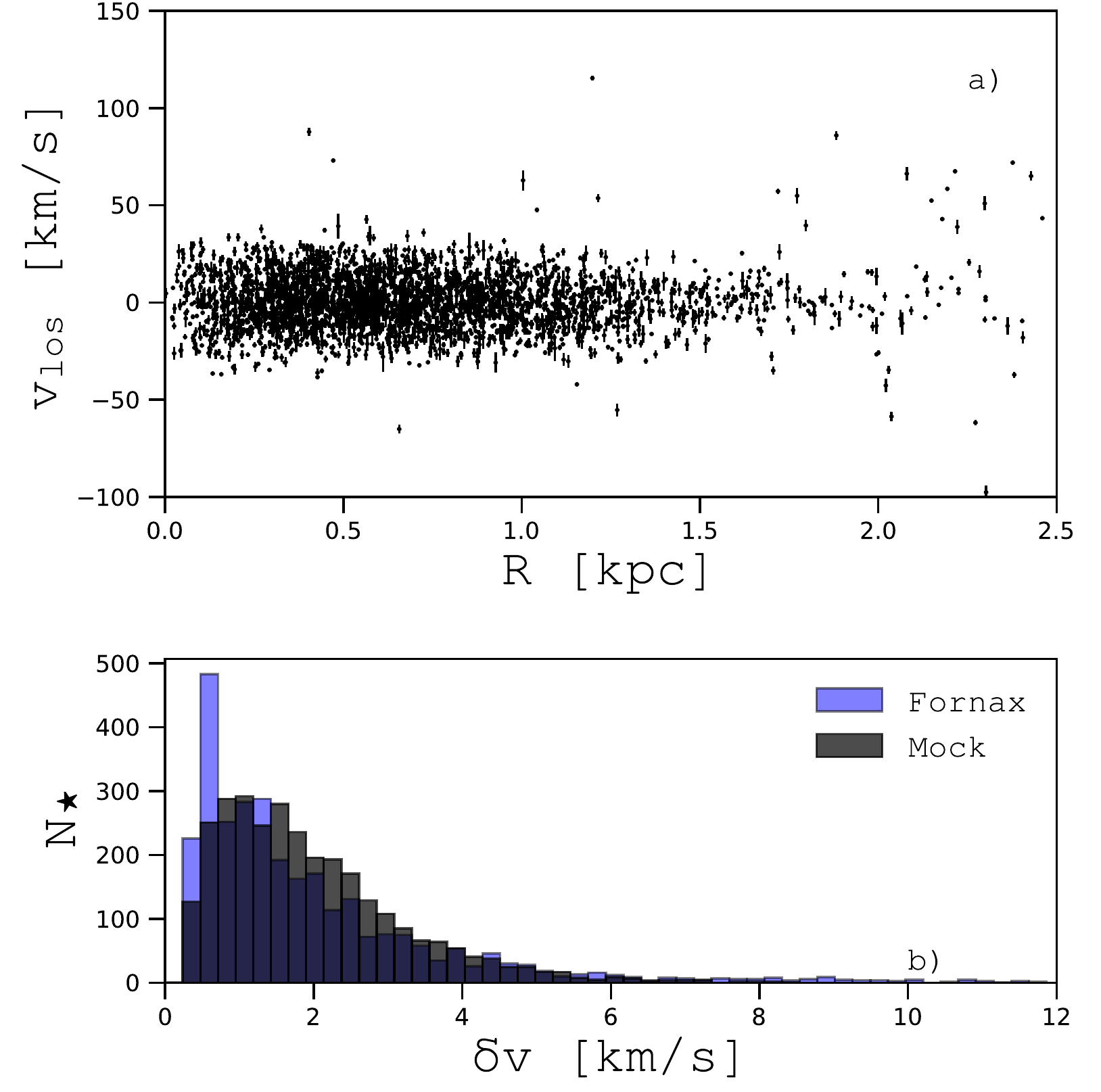}
 \caption{Panel a: position-velocity diagram of the mock kinematic sample. Panel b: mock error
 distribution (black histograms) superimposed to the error distribution of the Fornax kinematic
 sample (blue histograms).}\label{fig:mockerrors}
\end{figure}

We applied the $f({\bf J})$ models to a mock galaxy, with structure and kinematics
similar to a typical dSph such as Fornax, in order to test the accuracy of the method
presented in Section \ref{sec:Analysis2fj}.
The mock is an $N$-body representation of a spherically symmetric galaxy, embedded in an 
NFW-like dark halo. The density distribution of the stellar component is 

\begin{equation}\label{for:sersic_mock}
\rho_{\star}(r) =  \rho_S\biggl(\frac{r}{R_S}\biggr)^p \exp\biggl[-\biggl(\frac{r}{R_S}\biggr)^{\nu}\biggr],
\end{equation}
where $\rho_S$ and $R_S$ are, respectively, a reference density and a characteristic scale
radius, while $p = 1 - 0.6097\nu + 0.05463\nu^2$. Equation (\ref{for:sersic_mock}) is an
approximation to the deprojection of the \cite{Sersic1968} profile with index $m = 1/\nu$ 
(\citealt{Neto1999}), which usually gives a good representation of a dSph stellar surface 
density. The mock dark-matter density profile is 

\begin{equation}\label{for:dmcomp}
 \rho_{\rm dm} = \frac{\rho_0}{r/r_{\rm s, dm}(1 + r/r_{\rm s, dm})^2}e^{-\bigl(\frac{r}{r_{\rm t,dm}}\bigr)^2},
\end{equation}
where $\rho_0$ is a reference density, $r_{\rm s, dm}$ is the scale radius and 
$r_{\rm t, dm}$ is the truncation radius. Eddington's integral was used to compute the 
ergodic DF of the stellar component. 

The mock consists of 51200 stars. 
Each star was assigned position and velocity by using the DF. For the stellar component 
we used $m = 0.71$ and $R_S = 0.58\,$kpc (reference values of the S\'ersic best fit of the
Fornax projected number density profile; \citealt{Battaglia2006}), while for the 
dark-matter component $r_{\rm s, dm} = 4$, $r_{\rm t, dm} = 20$ and $\rho_0$ is such that the 
total dark matter mass $M_{\rm tot, dm} = 3\times10^9 M_{\odot}$. The total mass of the
mock is $\Mstot = 5\times10^7$ $M_{\odot}$. The parameters of the mock
are summarised in Table \ref{tab:mockParam}.

Using the terminology of Section \ref{sec:Analysis2fj}, we constructed the mock photometric and
kinematic sample (respectively, the projected surface density profile and the set of radial 
velocities with associated errors). For the mock we take a Cartesian system of coordinates such
that $(x,y)$ is the plane of the sky and $z$ is the line of sight. To these stars we added 
4700 stars with $(x,y)$ position randomly generated from a uniform two-dimensional distribution
and line-of-sight velocities from a normal LOSVD with mean 15 km s$^{-1}$ and standard deviation
40 km s$^{-1}$. The samples are computed as follows: 

\begin{itemize}
 \item[i)] {\em Photometric samples}. We divided the plane of the sky into four quadrants. The 
 projected number density profile has been computed in each quadrant, using $N_n = 50$ equally spaced 
 bins, centred at $R_i$, with $i=1,...,N_n$. For each bin we compute the mean $\mu_i$ and the standard 
 deviation $\sigma_i$ of the four measures. The projected number density profile is defined as
 $n^{\star}_i\equiv\mu_i$, with associated errors $\delta n^{\star}_i\equiv\sigma_i$. 
 From the outermost 23 bins, where the only contribution is that of contaminants, we evaluated 
 the background mean surface number density profile. Then, we take the first 27 bins and correct them 
 for the contamination.

 \item[ii)] {\em Kinematic sample}. From the whole mock we randomly selected $N_v = 3000$ stars,
 similar in size to the Fornax kinematic sample (Section \ref{sec:fornax}) with true velocities 
 along the z-axis $v_{z,i}$, with $k=1,...,N_v$.. 
 
 The distribution of line-of-sight velocity errors of the Fornax sample is skewed, with a 
 significant tail at large errors. To simulate the same effect, we randomly extracted the 
 errors on the mock velocities from a skewed beta distribution $B(a,b)$, with $a=1.5$ and $b=15$.
 The errors $\delta \vlosk$ have been scaled requiring that $\sigma_{\rm mock} / \delta 
 \overline{v}_{\rm mock} = 6.5$, where $\sigma_{\rm mock}$ is the standard deviation 
 of the radial velocity measurements and $\delta \overline{v}_{\rm mock}$ is the scaled mean
 of the velocity errors. Our final kinematic sample of mock velocities $\vlosk$ has been 
 computed by selecting randomly new velocities from a normal distribution with mean equal
 to $v_{z,i}$ and dispersion equal to the error on the k-th velocity $\delta v_{{\rm los}, k}$. 
\end{itemize}
Fig.~\ref{fig:mockerrors}a shows the position-velocity diagram of the mock, while 
Fig.~\ref{fig:mockerrors}b plots the error distribution of the mock kinematic sample
superimposed to the Fornax distribution of line-of-sight velocity errors.

We determined the model that best fits the mock applying the procedure described in Section \ref{sec:Analysis2fj}.
We analysed three different cases: a family with NFW-like halo and two families with cored halo,
with cores of different sizes. We will refer to the NFW family of models as mockNFW, and to the two cored
families as mockCore1 and mockCore2. The parameters of the best models are given in Table \ref{tab:mocktab}.
As for Fornax, we considered only models with total stellar mass in the range $\Mstot = 10^7 -10^8 M_{\odot}$
and halo scale radius in the range $2.5 \lesssim \rst \lesssim 6.2 $ (see Section \ref{subsec:insens}).

\begin{figure*}
 \centering
 \includegraphics[width=1\hsize]{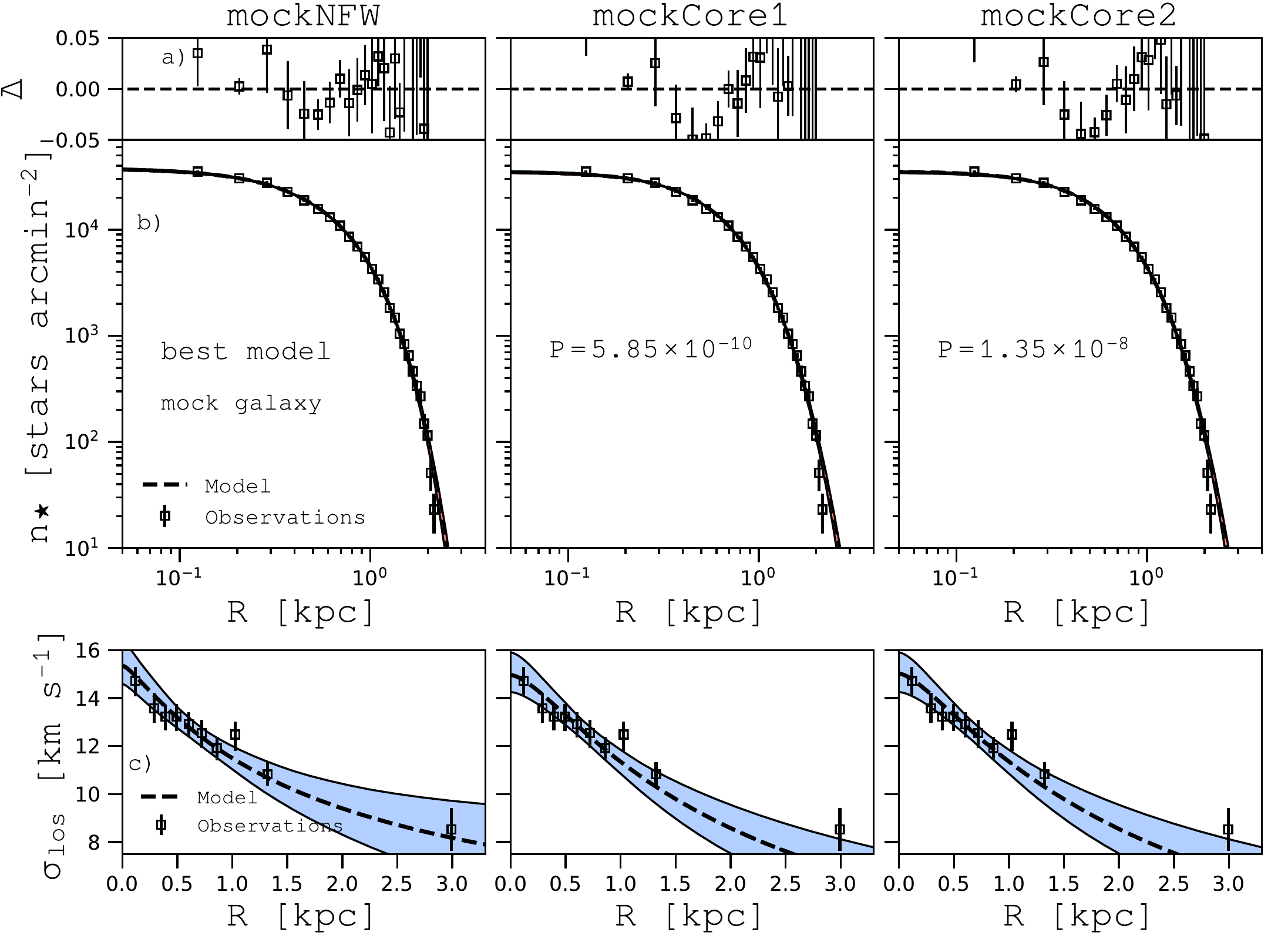}
 \caption{Columns, from left to right, refer to the best models of the mockNFW, mockCore1 and mockCore2
 families, respectively. Top row of panels (a): residuals $\Delta\equiv(\nobs-\nmod)/\nmod$ between the model 
 and the mock observed projected stellar density profile (points with error bars).
 Middle row of panels (b): projected number density profile of the model (dashed lines), compared with
 the mock observed profile (points with error bars). Bottom row of panels (c): line-of-sight 
 velocity dispersion profile of the model (dashed lines), compared  with the mock observed
 profile (points with error bars). In panels b and c the bands indicate the 1$\sigma$ 
 uncertainties (see Section \ref{subsec:models}).}\label{fig:dslosmock}
\end{figure*}

We were able to recover sufficiently well the total mass distribution of the mock galaxy: the cuspy halo is preferred with 
high significance over the two cored families here considered (see Table~\ref{tab:mocktab}). 
The projected stellar number density profiles and the line-of-sight velocity dispersion profiles of the best
models of the three families are compared with the corresponding profiles of the mock in Fig.~\ref{fig:dslosmock}. 
The line-of-sight velocity dispersion has been computed in 10 radial bins (each bin has 300 stars; see 
Section~\ref{subsec:twocb}). The best mockNFW model reproduces better than the best cored models both 
the projected number density profile and the line-of-sight velocity profile.

\begin{figure*}
 \centering
 \includegraphics[width=1\hsize]{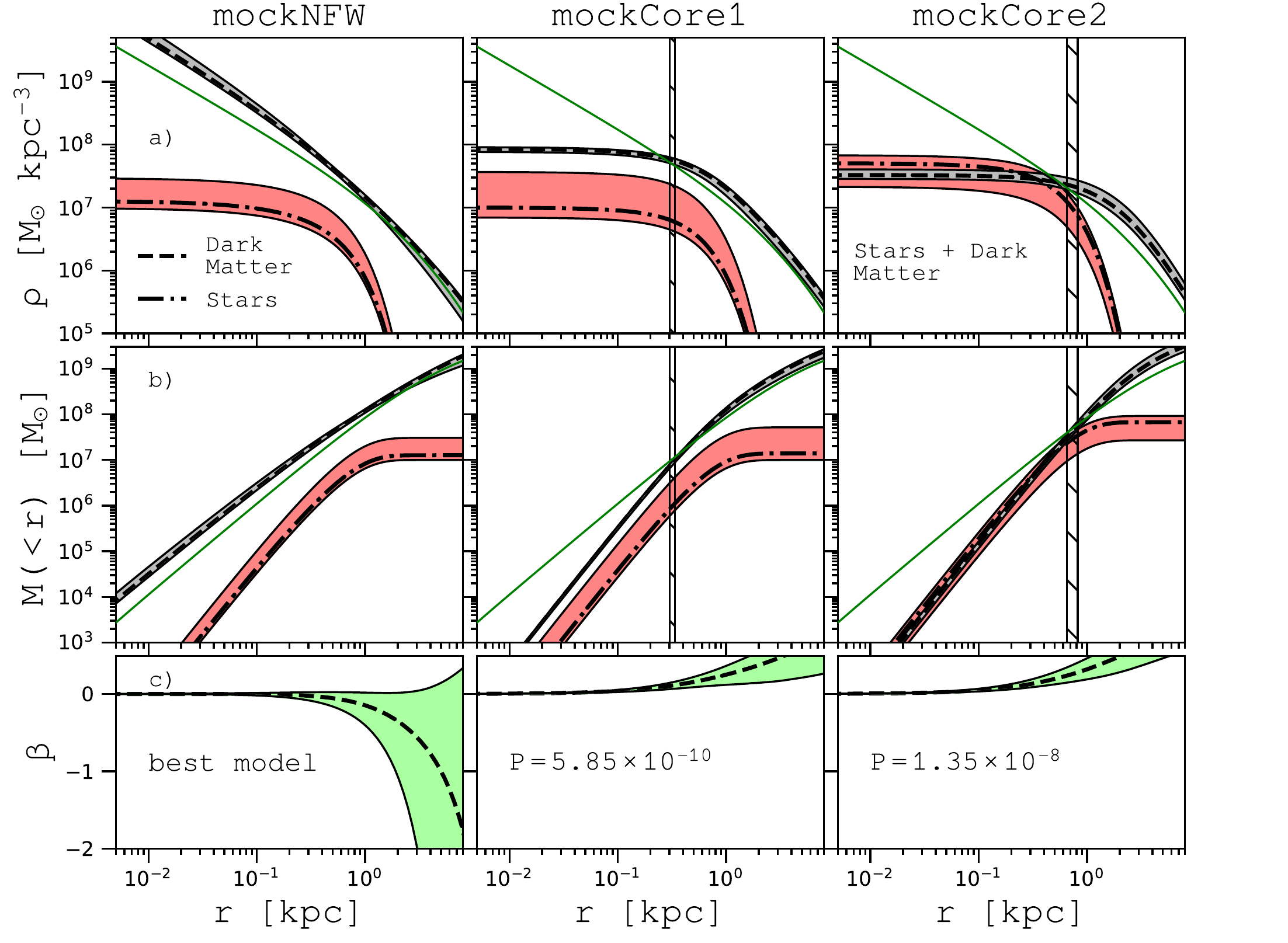}
 \caption{Columns, from left to right, refer to the best model of the mockNFW, mockCore1 and mockCore2
 families, respectively. Top row of panels (a): stellar (dash-dotted line) and dark-matter (dashed 
 line) density profiles. Middle row of panels (b): stellar (dash-dotted line) and dark-matter (dashed line)
 mass profiles. Bottom row of panels (c): anisotropy parameter profile (dashed line). The vertical black lines 
 mark the halo core radius $\rc$ for the cored families. In all
 panels the bands around the best fits indicate 1$\sigma$ uncertainty (see Section \ref{subsec:models}). 
 The green curve in panels (a) and (b) show, respectively, the density and mass distributions of the mock
 dark-matter component (equation \ref{for:dmcomp} and Table \ref{tab:mocktab}).}\label{fig:mock_d_m_a}
\end{figure*}

Fig.~\ref{fig:mock_d_m_a} plots the density, mass and anisotropy profiles of the best models of the three
families mockNFW, mockCore1 and mockCore2. The mock dark-matter mass distribution is well 
represented by the mockNFW best model. The differences between model and mock dark-matter density profiles
in the innermost regions are due to the fact that the DF (\ref{for:dmdf}) reproduces the asymptotic 
behavior of the analytic NFW profile, but not exactly its transition between the $\rho_{\rm dm}\sim 
r^{-1}$ and $\rho_{\rm dm}\sim r^{-3}$ regimes. We are not able to constrain the scale radius of the
dark halo $\rst$ (see Section \ref{subsec:insens}). We find that the best model has $\rst = 
5.98^{+0.22}_{-3.45}$, so all the explored values are within 1$\sigma$.
The anisotropy is well recovered within the $1\sigma$ uncertainty: we find that the best 
mockNFW family is consistent with the isotropic mock velocity distribution ($\beta = 0$)
over the entire radial range (Fig.~\ref{fig:mock_d_m_a}); the model anisotropy at 1 kpc is 
$\beta_{|1 \text{kpc}} = -0.15^{+0.16}_{-0.26}$.

Though the result of the application of our method to the mock is positive and reassuring, of course this
test is not meant to be a proof that our method would be able to recover the properties of any mock. For 
instance, we limited ourselves to the case of a system with isotropic velocity distribution and we considered 
only one realisation of the photometric and kinematic samples. However, to the extent that the explored mock is
an acceptable realisation of a dSph like Fornax, the result of our test suggests that if
Fornax had a cuspy dark halo our method should be able to detect it.

\bibliography{biblio}

\begin{thebibliography}{}
\makeatletter
\relax
\def\mn@urlcharsother{\let\do\@makeother \do\$\do\&\do\#\do\^\do\_\do\%\do\~}
\def\mn@doi{\begingroup\mn@urlcharsother \@ifnextchar [ {\mn@doi@}
  {\mn@doi@[]}}
\def\mn@doi@[#1]#2{\def\@tempa{#1}\ifx\@tempa\@empty \href
  {http://dx.doi.org/#2} {doi:#2}\else \href {http://dx.doi.org/#2} {#1}\fi
  \endgroup}
\def\mn@eprint#1#2{\mn@eprint@#1:#2::\@nil}
\def\mn@eprint@arXiv#1{\href {http://arxiv.org/abs/#1} {{\tt arXiv:#1}}}
\def\mn@eprint@dblp#1{\href {http://dblp.uni-trier.de/rec/bibtex/#1.xml}
  {dblp:#1}}
\def\mn@eprint@#1:#2:#3:#4\@nil{\def\@tempa {#1}\def\@tempb {#2}\def\@tempc
  {#3}\ifx \@tempc \@empty \let \@tempc \@tempb \let \@tempb \@tempa \fi \ifx
  \@tempb \@empty \def\@tempb {arXiv}\fi \@ifundefined
  {mn@eprint@\@tempb}{\@tempb:\@tempc}{\expandafter \expandafter \csname
  mn@eprint@\@tempb\endcsname \expandafter{\@tempc}}}

\bibitem[\protect\citeauthoryear{{Aaronson}}{{Aaronson}}{1983}]{Aaronson1986}
{Aaronson} M.,  1983, \mn@doi [\apjl] {10.1086/183969}, \href
  {http://adsabs.harvard.edu/abs/1983ApJ...266L..11A} {266, L11}

\bibitem[\protect\citeauthoryear{{Akaike}}{{Akaike}}{1998}]{Akaike1973}
{Akaike} H.,  1998.
Springer, New York, NY

\bibitem[\protect\citeauthoryear{{Amorisco} \& {Evans}}{{Amorisco} \&
  {Evans}}{2011}]{AmoriscoEvans2011}
{Amorisco} N.~C.,  {Evans} N.~W.,  2011, \mn@doi [\mnras]
  {10.1111/j.1365-2966.2010.17715.x}, \href
  {http://adsabs.harvard.edu/abs/2011MNRAS.411.2118A} {411, 2118}

\bibitem[\protect\citeauthoryear{{Amorisco}, {Agnello}  \& {Evans}}{{Amorisco}
  et~al.}{2013}]{Amorisco2013}
{Amorisco} N.~C.,  {Agnello} A.,   {Evans} N.~W.,  2013, \mn@doi [\mnras]
  {10.1093/mnrasl/sls031}, \href
  {http://adsabs.harvard.edu/abs/2013MNRAS.429L..89A} {429, L89}

\bibitem[\protect\citeauthoryear{{Arca-Sedda} \&
  {Capuzzo-Dolcetta}}{{Arca-Sedda} \& {Capuzzo-Dolcetta}}{2016}]{ArcaSedda2016}
{Arca-Sedda} M.,  {Capuzzo-Dolcetta} R.,  2016, \mn@doi [\mnras]
  {10.1093/mnras/stw1647}, \href
  {http://adsabs.harvard.edu/abs/2016MNRAS.461.4335A} {461, 4335}

\bibitem[\protect\citeauthoryear{{Arca-Sedda} \&
  {Capuzzo-Dolcetta}}{{Arca-Sedda} \& {Capuzzo-Dolcetta}}{2017}]{ArcaSedda2017}
{Arca-Sedda} M.,  {Capuzzo-Dolcetta} R.,  2017, \mn@doi [\mnras]
  {10.1093/mnras/stw2483}, \href
  {http://adsabs.harvard.edu/abs/2017MNRAS.464.3060A} {464, 3060}

\bibitem[\protect\citeauthoryear{{Battaglia} et~al.,}{{Battaglia}
  et~al.}{2006}]{Battaglia2006}
{Battaglia} G.,  et~al., 2006, \mn@doi [\aap] {10.1051/0004-6361:20065720},
  \href {http://adsabs.harvard.edu/abs/2006A%26A...459..423B} {459, 423}

\bibitem[\protect\citeauthoryear{{Battaglia}, {Helmi}, {Tolstoy}, {Irwin},
  {Hill}  \& {Jablonka}}{{Battaglia} et~al.}{2008}]{Battaglia2008}
{Battaglia} G.,  {Helmi} A.,  {Tolstoy} E.,  {Irwin} M.,  {Hill} V.,
  {Jablonka} P.,  2008, \mn@doi [\apjl] {10.1086/590179}, \href
  {http://adsabs.harvard.edu/abs/2008ApJ...681L..13B} {681, L13}

\bibitem[\protect\citeauthoryear{{Battaglia}, {Helmi}  \&
  {Breddels}}{{Battaglia} et~al.}{2013}]{Battaglia2013}
{Battaglia} G.,  {Helmi} A.,   {Breddels} M.,  2013, \mn@doi [\nar]
  {10.1016/j.newar.2013.05.003}, \href
  {http://adsabs.harvard.edu/abs/2013NewAR..57...52B} {57, 52}

\bibitem[\protect\citeauthoryear{{Battaglia}, {Sollima}  \&
  {Nipoti}}{{Battaglia} et~al.}{2015}]{Battaglia2015}
{Battaglia} G.,  {Sollima} A.,   {Nipoti} C.,  2015, \mn@doi [\mnras]
  {10.1093/mnras/stv2096}, \href
  {http://adsabs.harvard.edu/abs/2015MNRAS.454.2401B} {454, 2401}

\bibitem[\protect\citeauthoryear{{Binney}}{{Binney}}{2014}]{Binney2014}
{Binney} J.,  2014, \mn@doi [\mnras] {10.1093/mnras/stu297}, \href
  {http://adsabs.harvard.edu/abs/2014MNRAS.440..787B} {440, 787}

\bibitem[\protect\citeauthoryear{{Binney} \& {Piffl}}{{Binney} \&
  {Piffl}}{2015}]{Binney2015}
{Binney} J.,  {Piffl} T.,  2015, \mn@doi [\mnras] {10.1093/mnras/stv2225},
  \href {http://adsabs.harvard.edu/abs/2015MNRAS.454.3653B} {454, 3653}

\bibitem[\protect\citeauthoryear{{Breddels} \& {Helmi}}{{Breddels} \&
  {Helmi}}{2013}]{BreddelsHelmi2013}
{Breddels} M.~A.,  {Helmi} A.,  2013, \mn@doi [\aap]
  {10.1051/0004-6361/201321606}, \href
  {http://adsabs.harvard.edu/abs/2013A%26A...558A..35B} {558, A35}

\bibitem[\protect\citeauthoryear{{Bullock} \& {Boylan-Kolchin}}{{Bullock} \&
  {Boylan-Kolchin}}{2017}]{Bullock2017}
{Bullock} J.~S.,  {Boylan-Kolchin} M.,  2017, \mn@doi [\araa]
  {10.1146/annurev-astro-091916-055313}, \href
  {http://adsabs.harvard.edu/abs/2017ARA%26A..55..343B} {55, 343}

\bibitem[\protect\citeauthoryear{{Ciotti} \& {Morganti}}{{Ciotti} \&
  {Morganti}}{2010}]{CiottiMorganti2010}
{Ciotti} L.,  {Morganti} L.,  2010, \mn@doi [\mnras]
  {10.1111/j.1365-2966.2010.17184.x}, \href
  {http://adsabs.harvard.edu/abs/2010MNRAS.408.1070C} {408, 1070}

\bibitem[\protect\citeauthoryear{{Cole} \& {Binney}}{{Cole} \&
  {Binney}}{2017}]{ColeBinney2017}
{Cole} D.~R.,  {Binney} J.,  2017, \mn@doi [\mnras] {10.1093/mnras/stw2775},
  \href {http://adsabs.harvard.edu/abs/2017MNRAS.465..798C} {465, 798}

\bibitem[\protect\citeauthoryear{{Cole}, {Dehnen}  \& {Wilkinson}}{{Cole}
  et~al.}{2011}]{Cole2011}
{Cole} D.~R.,  {Dehnen} W.,   {Wilkinson} M.~I.,  2011, \mn@doi [\mnras]
  {10.1111/j.1365-2966.2011.19110.x}, \href
  {http://adsabs.harvard.edu/abs/2011MNRAS.416.1118C} {416, 1118}

\bibitem[\protect\citeauthoryear{{Das} \& {Binney}}{{Das} \&
  {Binney}}{2016}]{Das2016}
{Das} P.,  {Binney} J.,  2016, \mn@doi [\mnras] {10.1093/mnras/stw744}, \href
  {http://adsabs.harvard.edu/abs/2016MNRAS.460.1725D} {460, 1725}

\bibitem[\protect\citeauthoryear{{Diakogiannis}, {Lewis}, {Ibata}, {Guglielmo},
  {Kafle}, {Wilkinson}  \& {Power}}{{Diakogiannis}
  et~al.}{2017}]{Diakogiannis2017}
{Diakogiannis} F.~I.,  {Lewis} G.~F.,  {Ibata} R.~A.,  {Guglielmo} M.,  {Kafle}
  P.~R.,  {Wilkinson} M.~I.,   {Power} C.,  2017, \mn@doi [\mnras]
  {10.1093/mnras/stx1219}, \href
  {http://adsabs.harvard.edu/abs/2017MNRAS.470.2034D} {470, 2034}

\bibitem[\protect\citeauthoryear{{Einasto}}{{Einasto}}{1965}]{Einasto1965}
{Einasto} J.,  1965, Trudy Astrofizicheskogo Instituta Alma-Ata, \href
  {http://adsabs.harvard.edu/abs/1965TrAlm...5...87E} {5, 87}

\bibitem[\protect\citeauthoryear{{El-Zant}, {Shlosman}  \& {Hoffman}}{{El-Zant}
  et~al.}{2001}]{ElZant2001}
{El-Zant} A.,  {Shlosman} I.,   {Hoffman} Y.,  2001, \mn@doi [\apj]
  {10.1086/322516}, \href {http://adsabs.harvard.edu/abs/2001ApJ...560..636E}
  {560, 636}

\bibitem[\protect\citeauthoryear{{Evans}, {Sanders}  \&
  {Geringer-Sameth}}{{Evans} et~al.}{2016}]{Evans2016}
{Evans} N.~W.,  {Sanders} J.~L.,   {Geringer-Sameth} A.,  2016, \mn@doi [\prd]
  {10.1103/PhysRevD.93.103512}, \href
  {http://adsabs.harvard.edu/abs/2016PhRvD..93j3512E} {93, 103512}

\bibitem[\protect\citeauthoryear{{Goerdt}, {Moore}, {Read}, {Stadel}  \&
  {Zemp}}{{Goerdt} et~al.}{2006}]{Goerdt2006}
{Goerdt} T.,  {Moore} B.,  {Read} J.~I.,  {Stadel} J.,   {Zemp} M.,  2006,
  \mn@doi [\mnras] {10.1111/j.1365-2966.2006.10182.x}, \href
  {http://adsabs.harvard.edu/abs/2006MNRAS.368.1073G} {368, 1073}

\bibitem[\protect\citeauthoryear{{Goerdt}, {Moore}, {Read}  \&
  {Stadel}}{{Goerdt} et~al.}{2010}]{Goerdt2010}
{Goerdt} T.,  {Moore} B.,  {Read} J.~I.,   {Stadel} J.,  2010, \mn@doi [\apj]
  {10.1088/0004-637X/725/2/1707}, \href
  {http://adsabs.harvard.edu/abs/2010ApJ...725.1707G} {725, 1707}

\bibitem[\protect\citeauthoryear{{Hastings}}{{Hastings}}{1970}]{Hastings1970}
{Hastings} W.~K.,  1970, 57, 97

\bibitem[\protect\citeauthoryear{{Hui}, {Ostriker}, {Tremaine}  \&
  {Witten}}{{Hui} et~al.}{2017}]{Hui2017}
{Hui} L.,  {Ostriker} J.~P.,  {Tremaine} S.,   {Witten} E.,  2017, \mn@doi
  [\prd] {10.1103/PhysRevD.95.043541}, \href
  {http://adsabs.harvard.edu/abs/2017PhRvD..95d3541H} {95, 043541}

\bibitem[\protect\citeauthoryear{{Irwin} \& {Hatzidimitriou}}{{Irwin} \&
  {Hatzidimitriou}}{1995}]{IrwinHatzidimitriou1995}
{Irwin} M.,  {Hatzidimitriou} D.,  1995, \mn@doi [\mnras]
  {10.1093/mnras/277.4.1354}, \href
  {http://adsabs.harvard.edu/abs/1995MNRAS.277.1354I} {277, 1354}

\bibitem[\protect\citeauthoryear{{Jardel} \& {Gebhardt}}{{Jardel} \&
  {Gebhardt}}{2012}]{Jardel2012}
{Jardel} J.~R.,  {Gebhardt} K.,  2012, \mn@doi [\apj]
  {10.1088/0004-637X/746/1/89}, \href
  {http://adsabs.harvard.edu/abs/2012ApJ...746...89J} {746, 89}

\bibitem[\protect\citeauthoryear{{Jeffreson} et~al.,}{{Jeffreson}
  et~al.}{2017}]{Jeffreson2017}
{Jeffreson} S.~M.~R.,  et~al., 2017, \mn@doi [\mnras] {10.1093/mnras/stx1152},
  \href {http://adsabs.harvard.edu/abs/2017MNRAS.469.4740J} {469, 4740}

\bibitem[\protect\citeauthoryear{{King}}{{King}}{1966}]{King1966}
{King} I.~R.,  1966, \mn@doi [\aj] {10.1086/109857}, \href
  {http://adsabs.harvard.edu/abs/1966AJ.....71...64K} {71, 64}

\bibitem[\protect\citeauthoryear{{Kleyna}, {Wilkinson}, {Gilmore}  \&
  {Evans}}{{Kleyna} et~al.}{2003}]{Kleyna2003}
{Kleyna} J.~T.,  {Wilkinson} M.~I.,  {Gilmore} G.,   {Evans} N.~W.,  2003,
  \mn@doi [\apjl] {10.1086/375522}, \href
  {http://adsabs.harvard.edu/abs/2003ApJ...588L..21K} {588, L21}

\bibitem[\protect\citeauthoryear{{Lima Neto}, {Gerbal}  \& {M{\'a}rquez}}{{Lima
  Neto} et~al.}{1999}]{Neto1999}
{Lima Neto} G.~B.,  {Gerbal} D.,   {M{\'a}rquez} I.,  1999, \mn@doi [\mnras]
  {10.1046/j.1365-8711.1999.02849.x}, \href
  {http://adsabs.harvard.edu/abs/1999MNRAS.309..481L} {309, 481}

\bibitem[\protect\citeauthoryear{{Mashchenko}, {Couchman}  \&
  {Wadsley}}{{Mashchenko} et~al.}{2006}]{Mashchenko2006}
{Mashchenko} S.,  {Couchman} H.~M.~P.,   {Wadsley} J.,  2006, \mn@doi [\nat]
  {10.1038/nature04944}, \href
  {http://adsabs.harvard.edu/abs/2006Natur.442..539M} {442, 539}

\bibitem[\protect\citeauthoryear{{Mateo}}{{Mateo}}{1998}]{Mateo1998}
{Mateo} M.~L.,  1998, \mn@doi [\araa] {10.1146/annurev.astro.36.1.435}, \href
  {http://adsabs.harvard.edu/abs/1998ARA%26A..36..435M} {36, 435}

\bibitem[\protect\citeauthoryear{{Metropolis}, {Rosenbluth}, {Teller}  \&
  {Teller}}{{Metropolis} et~al.}{1953}]{Metropolis1953}
{Metropolis} A.~W.,  {Rosenbluth} M.~N.,  {Teller} A.~H.,   {Teller} E.,  1953,
  \mn@doi [Journal of Chemical Physics] {https://doi.org/10.1063/1.1699114},
  21, 1087

\bibitem[\protect\citeauthoryear{{Michie}}{{Michie}}{1963}]{Michie1963}
{Michie} R.~W.,  1963, \mn@doi [\mnras] {10.1093/mnras/126.6.499}, \href
  {http://adsabs.harvard.edu/abs/1963MNRAS.126..499M} {126, 499}

\bibitem[\protect\citeauthoryear{{Minor}, {Martinez}, {Bullock}, {Kaplinghat}
  \& {Trainor}}{{Minor} et~al.}{2010}]{Minor2010}
{Minor} Q.~E.,  {Martinez} G.,  {Bullock} J.,  {Kaplinghat} M.,   {Trainor} R.,
   2010, \mn@doi [\apj] {10.1088/0004-637X/721/2/1142}, \href
  {http://adsabs.harvard.edu/abs/2010ApJ...721.1142M} {721, 1142}

\bibitem[\protect\citeauthoryear{{Mo} \& {Mao}}{{Mo} \&
  {Mao}}{2004}]{MoMao2004}
{Mo} H.~J.,  {Mao} S.,  2004, \mn@doi [\mnras]
  {10.1111/j.1365-2966.2004.08114.x}, \href
  {http://adsabs.harvard.edu/abs/2004MNRAS.353..829M} {353, 829}

\bibitem[\protect\citeauthoryear{{Mu{\~n}oz-Cuartas}, {Macci{\`o}},
  {Gottl{\"o}ber}  \& {Dutton}}{{Mu{\~n}oz-Cuartas} et~al.}{2011}]{Cuartas2010}
{Mu{\~n}oz-Cuartas} J.~C.,  {Macci{\`o}} A.~V.,  {Gottl{\"o}ber} S.,   {Dutton}
  A.~A.,  2011, \mn@doi [\mnras] {10.1111/j.1365-2966.2010.17704.x}, \href
  {http://adsabs.harvard.edu/abs/2011MNRAS.411..584M} {411, 584}

\bibitem[\protect\citeauthoryear{{Navarro}, {Eke}  \& {Frenk}}{{Navarro}
  et~al.}{1996a}]{Navarro1996}
{Navarro} J.~F.,  {Eke} V.~R.,   {Frenk} C.~S.,  1996a, \mn@doi [\mnras]
  {10.1093/mnras/283.3.L72}, \href
  {http://adsabs.harvard.edu/abs/1996MNRAS.283L..72N} {283, L72}

\bibitem[\protect\citeauthoryear{{Navarro}, {Frenk}  \& {White}}{{Navarro}
  et~al.}{1996b}]{NavarroFrenkWhite1996}
{Navarro} J.~F.,  {Frenk} C.~S.,   {White} S.~D.~M.,  1996b, \mn@doi [\apj]
  {10.1086/177173}, \href {http://adsabs.harvard.edu/abs/1996ApJ...462..563N}
  {462, 563}

\bibitem[\protect\citeauthoryear{{Nipoti} \& {Binney}}{{Nipoti} \&
  {Binney}}{2015}]{NipotiBinney2015}
{Nipoti} C.,  {Binney} J.,  2015, \mn@doi [\mnras] {10.1093/mnras/stu2217},
  \href {http://adsabs.harvard.edu/abs/2015MNRAS.446.1820N} {446, 1820}

\bibitem[\protect\citeauthoryear{{Pe{\~n}arrubia}, {McConnachie}  \&
  {Navarro}}{{Pe{\~n}arrubia} et~al.}{2008}]{Penarrubia2008}
{Pe{\~n}arrubia} J.,  {McConnachie} A.~W.,   {Navarro} J.~F.,  2008, \mn@doi
  [\apj] {10.1086/521543}, \href
  {http://adsabs.harvard.edu/abs/2008ApJ...672..904P} {672, 904}

\bibitem[\protect\citeauthoryear{{Piffl}, {Penoyre}  \& {Binney}}{{Piffl}
  et~al.}{2015}]{Piffl2015}
{Piffl} T.,  {Penoyre} Z.,   {Binney} J.,  2015, \mn@doi [\mnras]
  {10.1093/mnras/stv938}, \href
  {http://adsabs.harvard.edu/abs/2015MNRAS.451..639P} {451, 639}

\bibitem[\protect\citeauthoryear{{Pontzen} \& {Governato}}{{Pontzen} \&
  {Governato}}{2012}]{Pontzen2012}
{Pontzen} A.,  {Governato} F.,  2012, \mn@doi [\mnras]
  {10.1111/j.1365-2966.2012.20571.x}, \href
  {http://adsabs.harvard.edu/abs/2012MNRAS.421.3464P} {421, 3464}

\bibitem[\protect\citeauthoryear{{Posti}, {Binney}, {Nipoti}  \&
  {Ciotti}}{{Posti} et~al.}{2015}]{Posti2015}
{Posti} L.,  {Binney} J.,  {Nipoti} C.,   {Ciotti} L.,  2015, \mn@doi [\mnras]
  {10.1093/mnras/stu2608}, \href
  {http://adsabs.harvard.edu/abs/2015MNRAS.447.3060P} {447, 3060}

\bibitem[\protect\citeauthoryear{{Pryor} \& {Meylan}}{{Pryor} \&
  {Meylan}}{1993}]{Pryor93}
{Pryor} C.,  {Meylan} G.,  1993, in {Djorgovski} S.~G.,  {Meylan} G.,  eds,
  Astronomical Society of the Pacific Conference Series Vol. 50, Structure and
  Dynamics of Globular Clusters. p.~357

\bibitem[\protect\citeauthoryear{{Read} \& {Gilmore}}{{Read} \&
  {Gilmore}}{2005}]{Read2005}
{Read} J.~I.,  {Gilmore} G.,  2005, \mn@doi [\mnras]
  {10.1111/j.1365-2966.2004.08424.x}, \href
  {http://adsabs.harvard.edu/abs/2005MNRAS.356..107R} {356, 107}

\bibitem[\protect\citeauthoryear{{Read}, {Iorio}, {Agertz}  \&
  {Fraternali}}{{Read} et~al.}{2017}]{Read2017}
{Read} J.~I.,  {Iorio} G.,  {Agertz} O.,   {Fraternali} F.,  2017, \mn@doi
  [\mnras] {10.1093/mnras/stx147}, \href
  {http://adsabs.harvard.edu/abs/2017MNRAS.467.2019R} {467, 2019}

\bibitem[\protect\citeauthoryear{{Richardson} \& {Fairbairn}}{{Richardson} \&
  {Fairbairn}}{2014}]{Richardson2014}
{Richardson} T.,  {Fairbairn} M.,  2014, \mn@doi [\mnras]
  {10.1093/mnras/stu691}, \href
  {http://adsabs.harvard.edu/abs/2014MNRAS.441.1584R} {441, 1584}

\bibitem[\protect\citeauthoryear{{Salucci}, {Wilkinson}, {Walker}, {Gilmore},
  {Grebel}, {Koch}, {Frigerio Martins}  \& {Wyse}}{{Salucci}
  et~al.}{2012}]{Salucci2012}
{Salucci} P.,  {Wilkinson} M.~I.,  {Walker} M.~G.,  {Gilmore} G.~F.,  {Grebel}
  E.~K.,  {Koch} A.,  {Frigerio Martins} C.,   {Wyse} R.~F.~G.,  2012, \mn@doi
  [\mnras] {10.1111/j.1365-2966.2011.20144.x}, \href
  {http://adsabs.harvard.edu/abs/2012MNRAS.420.2034S} {420, 2034}

\bibitem[\protect\citeauthoryear{{S\'anchez-Salcedo}, {Reyes-Iturbide}  \& J.{,
  Hernandez}}{{S\'anchez-Salcedo} et~al.}{2006}]{Sanchez2006}
{S\'anchez-Salcedo} F.~J.,  {Reyes-Iturbide}  J.{, Hernandez} X.,  2006,
  \mn@doi [\mnras] {10.1111/j.1365-2966.2006.10602.x}, \href
  {http://adsabs.harvard.edu/abs/2006MNRAS.370.1829S} {370, 1829}

\bibitem[\protect\citeauthoryear{{Sanders} \& {Binney}}{{Sanders} \&
  {Binney}}{2016}]{Sanders2016}
{Sanders} J.~L.,  {Binney} J.,  2016, \mn@doi [\mnras] {10.1093/mnras/stw106},
  \href {http://adsabs.harvard.edu/abs/2016MNRAS.457.2107S} {457, 2107}

\bibitem[\protect\citeauthoryear{{Sanders} \& {Evans}}{{Sanders} \&
  {Evans}}{2015}]{Sanders2015}
{Sanders} J.~L.,  {Evans} N.~W.,  2015, \mn@doi [\mnras]
  {10.1093/mnras/stv1898}, \href
  {http://adsabs.harvard.edu/abs/2015MNRAS.454..299S} {454, 299}

\bibitem[\protect\citeauthoryear{{Schwarzschild}}{{Schwarzschild}}{1979}]{Schwarzschild1979}
{Schwarzschild} M.,  1979, \mn@doi [\apj] {10.1086/157282}, \href
  {http://adsabs.harvard.edu/abs/1979ApJ...232..236S} {232, 236}

\bibitem[\protect\citeauthoryear{{Sersic}}{{Sersic}}{1968}]{Sersic1968}
{Sersic} J.~L.,  1968, {Atlas de galaxias australes}

\bibitem[\protect\citeauthoryear{{Shapley}}{{Shapley}}{1938}]{Shapley1938}
{Shapley} H.,  1938, \mn@doi [\nat] {10.1038/142715b0}, \href
  {http://adsabs.harvard.edu/abs/1938Natur.142..715S} {142, 715}

\bibitem[\protect\citeauthoryear{{Strigari}, {Bullock}, {Kaplinghat}, {Simon},
  {Geha}, {Willman}  \& {Walker}}{{Strigari} et~al.}{2008}]{Strigari2008}
{Strigari} L.~E.,  {Bullock} J.~S.,  {Kaplinghat} M.,  {Simon} J.~D.,  {Geha}
  M.,  {Willman} B.,   {Walker} M.~G.,  2008, \mn@doi [\nat]
  {10.1038/nature07222}, \href
  {http://adsabs.harvard.edu/abs/2008Natur.454.1096S} {454, 1096}

\bibitem[\protect\citeauthoryear{{Strigari}, {Frenk}  \& {White}}{{Strigari}
  et~al.}{2017}]{Strigari2017}
{Strigari} L.~E.,  {Frenk} C.~S.,   {White} S.~D.~M.,  2017, \mn@doi [\apj]
  {10.3847/1538-4357/aa5c8e}, \href
  {http://adsabs.harvard.edu/abs/2017ApJ...838..123S} {838, 123}

\bibitem[\protect\citeauthoryear{{Tollet} et~al.,}{{Tollet}
  et~al.}{2016}]{Tollet2016}
{Tollet} E.,  et~al., 2016, \mn@doi [\mnras] {10.1093/mnras/stv2856}, \href
  {http://adsabs.harvard.edu/abs/2016MNRAS.456.3542T} {456, 3542}

\bibitem[\protect\citeauthoryear{{Vasiliev}}{{Vasiliev}}{2018}]{Vasiliev2018}
{Vasiliev} E.,  2018, preprint, \href
  {http://adsabs.harvard.edu/abs/2018arXiv180208239V} {} (\mn@eprint {arXiv}
  {1802.08239})

\bibitem[\protect\citeauthoryear{{Walker} \& {Pe{\~n}arrubia}}{{Walker} \&
  {Pe{\~n}arrubia}}{2011}]{Walker2011}
{Walker} M.~G.,  {Pe{\~n}arrubia} J.,  2011, \mn@doi [\apj]
  {10.1088/0004-637X/742/1/20}, \href
  {http://adsabs.harvard.edu/abs/2011ApJ...742...20W} {742, 20}

\bibitem[\protect\citeauthoryear{{Walker}, {Mateo}  \& {Olszewski}}{{Walker}
  et~al.}{2009}]{Walker2009}
{Walker} M.~G.,  {Mateo} M.,   {Olszewski} E.~W.,  2009, \mn@doi [\aj]
  {10.1088/0004-6256/137/2/3100}, \href
  {http://adsabs.harvard.edu/abs/2009AJ....137.3100W} {137, 3100}

\bibitem[\protect\citeauthoryear{{Watkins}, {van de Ven}, {den Brok}  \& {van
  den Bosch}}{{Watkins} et~al.}{2013}]{Watkins2013}
{Watkins} L.~L.,  {van de Ven} G.,  {den Brok} M.,   {van den Bosch} R.~C.~E.,
  2013, \mn@doi [\mnras] {10.1093/mnras/stt1756}, \href
  {http://adsabs.harvard.edu/abs/2013MNRAS.436.2598W} {436, 2598}

\bibitem[\protect\citeauthoryear{{Williams} \& {Evans}}{{Williams} \&
  {Evans}}{2015}]{Williams2015}
{Williams} A.~A.,  {Evans} N.~W.,  2015, \mn@doi [\mnras]
  {10.1093/mnras/stv096}, \href
  {http://adsabs.harvard.edu/abs/2015MNRAS.448.1360W} {448, 1360}

\bibitem[\protect\citeauthoryear{{Wolf}, {Martinez}, {Bullock}, {Kaplinghat},
  {Geha}, {Mu{\~n}oz}, {Simon}  \& {Avedo}}{{Wolf} et~al.}{2010}]{Wolf2010}
{Wolf} J.,  {Martinez} G.~D.,  {Bullock} J.~S.,  {Kaplinghat} M.,  {Geha} M.,
  {Mu{\~n}oz} R.~R.,  {Simon} J.~D.,   {Avedo} F.~F.,  2010, \mn@doi [\mnras]
  {10.1111/j.1365-2966.2010.16753.x}, \href
  {http://adsabs.harvard.edu/abs/2010MNRAS.406.1220W} {406, 1220}

\bibitem[\protect\citeauthoryear{{Zhu}, {van de Ven}, {Watkins}  \&
  {Posti}}{{Zhu} et~al.}{2016}]{Zhu2016}
{Zhu} L.,  {van de Ven} G.,  {Watkins} L.~L.,   {Posti} L.,  2016, \mn@doi
  [\mnras] {10.1093/mnras/stw2081}, \href
  {http://adsabs.harvard.edu/abs/2016MNRAS.463.1117Z} {463, 1117}

\bibitem[\protect\citeauthoryear{{de Blok}}{{de Blok}}{2010}]{deBlok2009}
{de Blok} W.~J.~G.,  2010, \mn@doi [Advances in Astronomy]
  {10.1155/2010/789293}, \href
  {http://adsabs.harvard.edu/abs/2010AdAst2010E...5D} {2010, 789293}

\bibitem[\protect\citeauthoryear{{de Boer} et~al.,}{{de Boer}
  et~al.}{2012}]{deBoer2012}
{de Boer} T.~J.~L.,  et~al., 2012, \mn@doi [\aap]
  {10.1051/0004-6361/201219547}, \href
  {http://adsabs.harvard.edu/abs/2012A%26A...544A..73D} {544, A73}

\makeatother
\end{thebibliography}
\bibliographystyle{mnras}

\end{document}